\input harvmac

\noblackbox
\newdimen\tableauside\tableauside=1.0ex
\newdimen\tableaurule\tableaurule=0.4pt
\newdimen\tableaustep
\def\phantomhrule#1{\hbox{\vbox to0pt{\hrule height\tableaurule width#1\vss}}}
\def\phantomvrule#1{\vbox{\hbox to0pt{\vrule width\tableaurule height#1\hss}}}
\def\sqr{\vbox{%
  \phantomhrule\tableaustep
  \hbox{\phantomvrule\tableaustep\kern\tableaustep\phantomvrule\tableaustep}%
  \hbox{\vbox{\phantomhrule\tableauside}\kern-\tableaurule}}}
\def\squares#1{\hbox{\count0=#1\noindent\loop\sqr
  \advance\count0 by-1 \ifnum\count0>0\repeat}}
\def\tableau#1{\vcenter{\offinterlineskip
  \tableaustep=\tableauside\advance\tableaustep by-\tableaurule
  \kern\normallineskip\hbox
    {\kern\normallineskip\vbox
      {\gettableau#1 0 }%
     \kern\normallineskip\kern\tableaurule}%
  \kern\normallineskip\kern\tableaurule}}
\def\gettableau#1 {\ifnum#1=0\let\next=\null\else
  \squares{#1}\let\next=\gettableau\fi\next}

\tableauside=1.0ex
\tableaurule=0.4pt
\input epsf
\noblackbox

\def\CN{{\cal N}}



\def\unlockat{\catcode`\@=11}
\def\lockat{\catcode`\@=12}

\unlockat

\def\newsec#1{\global\advance\secno by1\message{(\the\secno. #1)}
\global\subsecno=0\global\subsubsecno=0\eqnres@t\noindent
{\bf\the\secno. #1}
\writetoca{{\secsym} {#1}}\par\nobreak\medskip\nobreak}
\global\newcount\subsecno \global\subsecno=0
\def\subsec#1{\global\advance\subsecno
by1\message{(\secsym\the\subsecno. #1)}
\ifnum\lastpenalty>9000\else\bigbreak\fi\global\subsubsecno=0
\noindent{\it\secsym\the\subsecno. #1}
\writetoca{\string\quad {\secsym\the\subsecno.} {#1}}
\par\nobreak\medskip\nobreak}
\global\newcount\subsubsecno \global\subsubsecno=0
\def\subsubsec#1{\global\advance\subsubsecno
\message{(\secsym\the\subsecno.\the\subsubsecno. #1)}
\ifnum\lastpenalty>9000\else\bigbreak\fi
\noindent\quad{\secsym\the\subsecno.\the\subsubsecno.}{#1}
\writetoca{\string\qquad{\secsym\the\subsecno.\the\subsubsecno.}{#1}}
\par\nobreak\medskip\nobreak}

\def\subsubseclab#1{\DefWarn#1\xdef
#1{\noexpand\hyperref{}{subsubsection}%
{\secsym\the\subsecno.\the\subsubsecno}%
{\secsym\the\subsecno.\the\subsubsecno}}%
\writedef{#1\leftbracket#1}\wrlabeL{#1=#1}}
\lockat

\def\IL{{\relax{\rm I\kern-.18em L}}}
\def\IH{{\relax{\rm I\kern-.18em H}}}
\def\IR{{\relax{\rm I\kern-.18em R}}}
\def\IE{{\relax{\rm I\kern-.18em E}}}
\def\IC{{\relax\hbox{$\inbar\kern-.3em{\rm C}$}}}
\def\IZ{{\relax\ifmmode\mathchoice
{\hbox{\cmss Z\kern-.4em Z}}{\hbox{\cmss Z\kern-.4em Z}}
{\lower.9pt\hbox{\cmsss Z\kern-.4em Z}}
{\lower1.2pt\hbox{\cmsss Z\kern-.4em Z}}\else{\cmss Z\kern-.4em
Z}\fi}}
\def\CM {{\cal M}}
\def\CN {{\cal N}}

\def\CD {{\cal D}}

\def\CP {{\cal P }}
\def\CL {{\cal L}}

\def\CO {{\cal O}}

\def\CC {{\cal C}}

\def\CS {{\cal S}}

\def\CM {{\cal M}}
\def\CN {{\cal N}}

\def\CO {{\cal O}}

\def\CP {{\cal P }}

\def\CS {{\cal S }}
\def\CT{{\cal T}}
\def\ch{{\rm ch}}

\font\manual=manfnt \def\dbend{\lower3.5pt\hbox{\manual\char127}}

\def\IZ{{\relax\ifmmode\mathchoice
{\hbox{\cmss Z\kern-.4em Z}}{\hbox{\cmss Z\kern-.4em Z}}
{\lower.9pt\hbox{\cmsss Z\kern-.4em Z}}
{\lower1.2pt\hbox{\cmsss Z\kern-.4em Z}}\else{\cmss Z\kern-.4em
Z}\fi}}
\def\half {{1\over 2}}

\def\CM {{\cal M}}
\def\CN {{\cal N}}

\def\CO {{\cal O}}

\def\CP {{\cal P }}

\def\CS {{\cal S }}

\def\ch{{\rm ch}}


\def\IZ{{\relax\ifmmode\mathchoice
{\hbox{\cmss Z\kern-.4em Z}}{\hbox{\cmss Z\kern-.4em Z}}
{\lower.9pt\hbox{\cmsss Z\kern-.4em Z}}
{\lower1.2pt\hbox{\cmsss Z\kern-.4em Z}}\else{\cmss Z\kern-.4em
Z}\fi}}
\def\IB{{\relax{\rm I\kern-.18em B}}}
\def\IC{{\relax\hbox{$\inbar\kern-.3em{\rm C}$}}}
\def\ID{{\relax{\rm I\kern-.18em D}}}
\def\IE{{\relax{\rm I\kern-.18em E}}}
\def\IF{{\relax{\rm I\kern-.18em F}}}
\def\IG{{\relax\hbox{$\inbar\kern-.3em{\rm G}$}}}
\def\IGa{{\relax\hbox{${\rm I}\kern-.18em\Gamma$}}}
\def\IH{{\relax{\rm I\kern-.18em H}}}
\def\II{{\relax{\rm I\kern-.18em I}}}
\def\IK{{\relax{\rm I\kern-.18em K}}}
\def\IP{{\relax{\rm I\kern-.18em P}}}

\def\inbar{\,\vrule height1.5ex width.4pt depth0pt}

\font\cmss=cmss10 \font\cmsss=cmss10 at 7pt
\def\IR{\relax{\rm I\kern-.18em R}}
\def\IT{\relax{\rm I\kern-.18em T}}


\def\boxit#1{\vbox{\hrule\hbox{\vrule\kern8pt
\vbox{\hbox{\kern8pt}\hbox{\vbox{#1}}\hbox{\kern8pt}}
\kern8pt\vrule}\hrule}}
\def\mathboxit#1{\vbox{\hrule\hbox{\vrule\kern8pt\vbox{\kern8pt
\hbox{$\displaystyle #1$}\kern8pt}\kern8pt\vrule}\hrule}}


\def\inbar{\,\vrule height1.5ex width.4pt depth0pt}

\font\cmss=cmss10 \font\cmsss=cmss10 at 7pt
\def\IR{\relax{\rm I\kern-.18em R}}


\def\half{{1\over 2}}

\def\exp{{\rm exp}}

\def\wSigma{{\widetilde \Sigma}}
\def\wP{{\widetilde P}}
\let\includefigures=\iftrue
\newfam\black
\includefigures

\input epsf
\def\plb#1 #2 {Phys. Lett. {\bf B#1} #2 }
\long\def\del#1\enddel{}
\long\def\new#1\endnew{{\bf #1}}
\let\<\langle \let\>\rangle

\def\figin{\epsfcheck\figin}\def\figins{\epsfcheck\figins}
\def\epsfcheck{\ifx\epsfbox\UnDeFiNeD
\message{(NO epsf.tex, FIGURES WILL BE IGNORED)}
\gdef\figin##1{\vskip2in}\gdef\figins##1{\hskip.5in} blank space instead
\else\message{(FIGURES WILL BE INCLUDED)}
\gdef\figin##1{##1}\gdef\figins##1{##1}\fi}
\def\DefWarn#1{}
\def\figinsert{\goodbreak\midinsert}
\def\ifig#1#2#3{\DefWarn#1\xdef#1{fig.~\the\figno}
\writedef{#1\leftbracket fig.\noexpand~\the\figno}
\figinsert\figin{\centerline{#3}}\medskip
\centerline{\vbox{\baselineskip12pt
\advance\hsize by -1truein\noindent
\footnotefont{\bf Fig.~\the\figno:} #2}}
\bigskip\endinsert\global\advance\figno by1}
\else
\def\ifig#1#2#3{\xdef#1{fig.~\the\figno}
\writedef{#1\leftbracket fig.\noexpand~\the\figno}
\figinsert\figin{\centerline{#3}}\medskip
\centerline{\vbox{\baselineskip12pt
\advance\hsize by -1truein\noindent
\footnotefont{\bf Fig.~\the\figno:} #2}}
\bigskip\endinsert
\global\advance\figno by1}
\fi

\input xy
\xyoption{all}
\font\cmss=cmss10 \font\cmsss=cmss10 at 7pt
\def\inbar{\,\vrule height1.5ex width.4pt depth0pt}
\def\IC{{\relax\hbox{$\inbar\kern-.3em{\rm C}$}}}
\def\IP{{\relax{\rm I\kern-.18em P}}}
\def\IF{{\relax{\rm I\kern-.18em F}}}
\def\IZ{\relax\ifmmode\mathchoice
{\hbox{\cmss Z\kern-.4em Z}}{\hbox{\cmss Z\kern-.4em Z}}
{\lower.9pt\hbox{\cmsss Z\kern-.4em Z}}
{\lower1.2pt\hbox{\cmsss Z\kern-.4em Z}}\else{\cmss Z\kern-.4em
Z}\fi}
\def\IR{{\relax{\rm I\kern-.18em R}}}
\def\IQ{\relax\hbox{\kern.25em$\inbar\kern-.3em{\rm Q}$}}

\def\pmb#1{\setbox0=\hbox{#1}%
 \kern-.025em\copy0\kern-\wd0
 \kern.05em\copy0\kern-\wd0
 \kern-.025em\raise.0433em\box0 }
\font\cmss=cmss10
\font\cmsss=cmss10 at 7pt
\def\rlx{\relax\leavevmode}
\def\Cop{\relax\,\hbox{$\inbar\kern-.3em{\rm C}$}}
\def\Rop{\relax{\rm I\kern-.18em R}}
\def\Nop{\relax{\rm I\kern-.18em N}}
\def\Pop{\relax{\rm I\kern-.18em P}}
\def\Zop{\rlx\leavevmode\ifmmode\mathchoice{\hbox{\cmss Z\kern-.4em Z}}
 {\hbox{\cmss Z\kern-.4em Z}}{\lower.9pt\hbox{\cmsss Z\kern-.36em Z}}
 {\lower1.2pt\hbox{\cmsss Z\kern-.36em Z}}\else{\cmss Z\kern-.4em
 Z}\fi}

\def\inbar{\,\vrule height1.5ex width.4pt depth0pt}
\def\IC{{\relax\hbox{$\inbar\kern-.3em{\rm C}$}}}
\def\IP{{\relax{\rm I\kern-.18em P}}}
\def\IF{{\relax{\rm I\kern-.18em F}}}
\def\IZ{\relax\ifmmode\mathchoice
{\hbox{\cmss Z\kern-.4em Z}}{\hbox{\cmss Z\kern-.4em Z}}
{\lower.9pt\hbox{\cmsss Z\kern-.4em Z}}
{\lower1.2pt\hbox{\cmsss Z\kern-.4em Z}}\else{\cmss Z\kern-.4em
Z}\fi}
\def\IR{{\relax{\rm I\kern-.18em R}}}

\def\half{{1\over 2}}

\def\GeV{{\rm GeV}}

\def\IL{{\relax{\rm I\kern-.18em L}}}
\def\IH{{\relax{\rm I\kern-.18em H}}}
\def\IR{{\relax{\rm I\kern-.18em R}}}
\def\IE{{\relax{\rm I\kern-.18em E}}}
\def\IC{{\relax\hbox{$\inbar\kern-.3em{\rm C}$}}}
\def\IZ{{\relax\ifmmode\mathchoice
{\hbox{\cmss Z\kern-.4em Z}}{\hbox{\cmss Z\kern-.4em Z}}
{\lower.9pt\hbox{\cmsss Z\kern-.4em Z}}
{\lower1.2pt\hbox{\cmsss Z\kern-.4em Z}}\else{\cmss Z\kern-.4em
Z}\fi}}

\def\CM {{\cal M}}
\def\CN {{\cal N}}

\def\CD {{\cal D}}

\def\CP {{\cal P }}
\def\CL {{\cal L}}

\def\CO {{\cal O}}

\def\CC {{\cal C}}

\def\CS {{\cal S}}


\def\CM {{\cal M}}
\def\CN {{\cal N}}

\def\CO {{\cal O}}

\def\CP {{\cal P }}

\def\CS {{\cal S }}
\def\CT{{\cal T}}
\def\ch{{\rm ch}}

\def\wS{{\widetilde S}}
\def\wP{{\widetilde P}}
\def\kah{{K\"ahler}}
\font\manual=manfnt \def\dbend{\lower3.5pt\hbox{\manual\char127}}

\def\IZ{{\relax\ifmmode\mathchoice
{\hbox{\cmss Z\kern-.4em Z}}{\hbox{\cmss Z\kern-.4em Z}}
{\lower.9pt\hbox{\cmsss Z\kern-.4em Z}}
{\lower1.2pt\hbox{\cmsss Z\kern-.4em Z}}\else{\cmss Z\kern-.4em
Z}\fi}}
\def\half {{1\over 2}}

\def\CM {{\cal M}}
\def\CN {{\cal N}}

\def\CO {{\cal O}}

\def\CP {{\cal P }}

\def\CS {{\cal S }}

\def\ch{{\rm ch}}


\def\IZ{{\relax\ifmmode\mathchoice
{\hbox{\cmss Z\kern-.4em Z}}{\hbox{\cmss Z\kern-.4em Z}}
{\lower.9pt\hbox{\cmsss Z\kern-.4em Z}}
{\lower1.2pt\hbox{\cmsss Z\kern-.4em Z}}\else{\cmss Z\kern-.4em
Z}\fi}}
\def\IB{{\relax{\rm I\kern-.18em B}}}
\def\IC{{\relax\hbox{$\inbar\kern-.3em{\rm C}$}}}
\def\ID{{\relax{\rm I\kern-.18em D}}}
\def\IE{{\relax{\rm I\kern-.18em E}}}
\def\IF{{\relax{\rm I\kern-.18em F}}}
\def\IG{{\relax\hbox{$\inbar\kern-.3em{\rm G}$}}}
\def\IGa{{\relax\hbox{${\rm I}\kern-.18em\Gamma$}}}
\def\IH{{\relax{\rm I\kern-.18em H}}}
\def\II{{\relax{\rm I\kern-.18em I}}}
\def\IK{{\relax{\rm I\kern-.18em K}}}
\def\IP{{\relax{\rm I\kern-.18em P}}}

\def\inbar{\,\vrule height1.5ex width.4pt depth0pt}

\lref\GR{G. F. Giudice and R. Rattazzi, ``Theories with Gauge Mediated
Supersymmetry Breaking,'' Phys. Rept. {\bf 322} (1999) 419, hep-ph/9801271.}
\lref\Dreview{For a review with further references, see:
R. Blumenhagen, M. Cvetic, P. Langacker and G. Shiu,
``Toward Realistic Intersecting D-Brane Models,'' hep-th/0502005.}
\lref\HKR{L. Hall, V. Kostelecky and S. Raby, ``New Flavor Violations
in Supergravity Models,'' Nucl. Phys. {\bf B267} (1986) 415.}
\lref\quiverDSB{D. Berenstein, C. Herzog, P. Ouyang and S. Pinansky,
``Supersymmetry Breaking from a Calabi-Yau Singularity,'' hep-th/0505029;
S. Franco, A. Hanany, F. Saad and A. Uranga, ``Fractional Branes and
Dynamical Supersymmetry Breaking,'' hep-th/0505040;
M. Bertolini, F. Bigazzi and A. Cotrone, ``Supersymmetry Breaking at
the End of a Cascade of Seiberg Dualities,'' hep-th/0505055.}
\lref\BP{
R. Bousso and J. Polchinski, ``Quantization of Four-Form Fluxes and
Dynamical Neutralization of the Cosmological Constant,''
JHEP {\bf 0006} (2000) 006, hep-th/0004134.}
\lref\DM{M. R. Douglas and G. Moore, ``D-Branes, Quivers and ALE Instantons,''
hep-th/9603167.}
\lref\Douglas{See for instance:
M. R. Douglas, ``Basic Results in Vacuum Statistics,''
Comptes Rendus Physique {\bf 5} (2004) 965, hep-th/0409207.}

\lref\orbCFT{S. Kachru and E. Silverstein, ``$4D$ Conformal Field Theories
and Strings on Orbifolds,'' Phys. Rev. Lett. {\bf 80} (1998) 4855,
hep-th/9802183.}

\lref\orbtwo{A. Lawrence, N. Nekrasov and C. Vafa, ``On Conformal Field
Theories in Four Dimensions,'' hep-th/9803015.}

\lref\SVW{S. Sethi, C. Vafa and E. Witten, ``Constraints on Low-Dimensional
String Compactifications,'' Nucl. Phys. {\bf B480} 213 (1996), hep-th/9606122.}

\lref\EvaAlex{A. Saltman and E. Silverstein, ``The Scaling of the No
Scale Potential and de Sitter Model Building,'' JHEP {\bf 0411} (2004)
066, hep-th/0402135; ``A New Handle on de Sitter Compactifications,''
hep-th/0411271.}

\lref\KW{I. Klebanov and E. Witten, ``Superconformal Field Theory
on Threebranes at a Calabi-Yau Singularity,'' Nucl. Phys. {\bf B536} (1998)
199, hep-th/9807080.} 

\lref\DGT{M. Dine, E. Gorbatov and S. Thomas, ``Low-Energy Supersymmetry
from the Landscape,'' hep-th/0407043.}

\lref\KPV{S. Kachru, J. Pearson and H. Verlinde, ``Brane/Flux Annihilaton
and the String Dual of a Non-Supersymmetric Field Theory,''
JHEP {\bf 0206} (2002) 021, hep-th/0112197.}

\lref\Dougstat{M. R. Douglas, ``The Statistics of String/M-Theory Vacua,''
JHEP {\bf 0305} (2003) 046, hep-th/0303194.}

\lref\Dine{M. Dine, ``The Intermediate Scale Branch of the Landscape,''
hep-th/0505202.}

\lref\KS{I. Klebanov and M. Strassler, ``Supergravity and a Confining
Gauge Theory: Duality Cascades and $\chi$SB Resolution of Naked
Singularities,'' JHEP {\bf 0008} (2000) 052, hep-th/0007191.}

\lref\DKS{A. Dymarsky, I. Klebanov and N. Seiberg, ``On the Moduli Space
of the Cascading $SU(M+P) \times SU(P)$ Gauge Theory,'' hep-th/0511254.}

\lref\GM{Z. Chacko, M. Luty, A. Nelson and E. Ponton, ``Gaugino
Mediated Supersymmetry Breaking,'' JHEP {\bf 0001} (2000) 003,
hep-ph/9911323;
D. Kaplan, G. Kribs and M. Schmaltz, ``Supersymmetry Breaking Through
Transparent Extra Dimensions,'' Phys. Rev. {\bf D62} (2000) 035010,
hep-ph/9911293.}

\lref\Anom{L. Randall and R. Sundrum, ``Out of This World
Supersymmetry Breaking,'' Nucl. Phys. {\bf B557} (1999) 79,
hep-th/9810155;
G. Giudice, M. Luty, H. Murayama and R. Rattazzi, ``Gaugino Mass
Without Singlets,'' JHEP {\bf 9812} (1998) 027, hep-ph/9810442.}

\lref\softsusy{
B. Allanach, F. Quevedo and K. Suruliz, ``Low-Energy Supersymmetry Breaking
from String Flux Compactifications: Benchmark Scenarios,'' hep-ph/0512081;
K. Choi, ``Moduli Stabilization and the Pattern of Soft SUSY Breaking
Terms,'' hep-ph/0511162;
J. Conlon, F. Quevedo and K. Suruliz, ``Large Volume Flux Compactifications:
Moduli Spectrum and D3/D7 Soft Supersymmetry Breaking,'' JHEP {\bf 0508}
(2005) 007, hep-th/0505076;
K. Choi, A. Falkowski, H. Nilles and M. Olechowski, ``Soft Supersymmetry
Breaking in KKLT Flux Compactification,'' hep-th/0503216;
D. L\"ust, P. Mayr, S. Reffert and S. Stieberger, ``F-Theory Flux,
Destabilization of Orientifolds and Soft Terms on D7-Branes,'' hep-th/0501139;
F. Marchesano, G. Shiu and L. Wang, ``Model Building and
Phenomenology of Flux-Induced Supersymmetry Breaking on
D3-Branes,'' Nucl. Phys. {\bf B712} (2005) 20, hep-th/0411080;
D. L\"ust, S. Reffert and S. Stieberger, ``MSSM with Soft Susy
Breaking Terms from D7-Branes,'' hep-th/0410074;
L. Ibanez, ``The Fluxed MSSM,'' Phys. Rev. {\bf D71} (2005) 055005,
hep-ph/0408064;
P. Camara, L. Ibanez and A. Uranga, ``Flux-Induced Susy Breaking
Soft Terms on D7-D3 Brane Systems,'' Nucl. Phys. {\bf B708} (2005) 268,
hep-th/0408036;
D. L\"ust, S. Reffert and S. Stieberger, ``Flux-Induced Soft Supersymmetry
Breaking in Chiral Type IIB Orientifolds with D3/D7-Branes,''
Nucl. Phys. {\bf B706} (2005) 3, hep-th/0406092;
A. Lawrence and J. McGreevy, ``Local String Models of Soft
Supersymmetry Breaking,'' JHEP {\bf 0406} (2004) 007, hep-th/0401034;
M. Grana, T. Grimm, H. Jockers and J. Louis, ``Soft Supersymmetry
Breaking in Calabi-Yau Orientifolds with D-Branes and Fluxes,''
Nucl. Phys. {\bf B690} (2004) 21, hep-th/0312232;
M. Grana, ``MSSM parameters from supergravity backgrounds,'' Phys.
Rev. {\bf D67} (2003) 066006, hep-th/0209200.
}

\lref\Herman{H. Verlinde, ``Holography and Compactification,''
Nucl. Phys. {\bf B580} (2000) 264, hep-th/9906182.}

\lref\Edinst{E. Witten, ``Nonperturbative Superpotentials in String Theory,''
Nucl. Phys. {\bf B474} (1996) 343, hep-th/9604030.}

\lref\GKP{S. B. Giddings, S. Kachru and J. Polchinski, ``Hierarchies
from Fluxes in String Compactifications,'' Phys. Rev. {\bf D66} (2002) 106006,
hep-th/0105097.}

\lref\GKTT{L. G\"orlich, S. Kachru, P. Tripathy and S. Trivedi, ``Gaugino
Condensation and Nonperturbative Superpotentials in Flux Compactifications,''
JHEP {\bf 0412} (2004) 074, hep-th/0407130;
J. Cascales and A. Uranga, ``Branes on Generalized Calibrated Submanifolds,''
JHEP {\bf 0411} (2004) 083, hep-th/0407132.}

\lref\HW{P. Horava and E. Witten, ``Heterotic and Type I String
Dynamics from Eleven Dimensions,'' Nucl. Phys. {\bf B460} (1996) 506,
hep-th/9510209.}

\lref\WittGG{E. Witten, ``Strong Coupling Expansion of Calabi-Yau
Compactification,'' Nucl. Phys. {\bf B471} (1996) 135, hep-th/9602070.}

\lref\unification{S. Dimopoulos, S. Raby and F. Wilczek, ``Supersymmetry
and the Scale of Unification,'' Phys. Rev. {\bf D24} (1981) 1681.}

\lref\MSSM{S. Dimopoulos and H. Georgi, ``Softly Broken Supersymmetry
and $SU(5)$," Nucl. Phys. {\bf B193} (1981) 150.}

\lref\Nilles{H. Nilles, ``Supersymmetry, Supergravity and Particle
Physics,'' Phys. Rept. {\bf 110} (1984) 1.}

\lref\WittenDSB{E. Witten, ``Dynamical Breaking of Supersymmetry,''
Nucl. Phys. {\bf B188} (1981) 513.}

\lref\Verlinde{H. Verlinde and M. Wijnholt, ``Building the Standard
Model on a D3-Brane,'' hep-th/0508089.}

\lref\bottomup{G. Aldazabal, L. Ibanez, F. Quevedo and A. Uranga,
``D-Branes at Singularities: a Bottom-up Approach to the String
Embedding of the Standard Model,'' JHEP {\bf 0008} (2000) 002, hep-th/0005067;
R. Blumenhagen, B. K\"ors, D. L\"ust and T. Ott, ``The Standard Model
from Stable Intersecting Brane World Orbifolds,'' Nucl. Phys.
{\bf B616} (2001) 3, hep-th/0107138;
M. Cvetic, G. Shiu and A. Uranga, ``Three Family Supersymmetric Standard-like
Models from Intersecting Brane Worlds,'' Phys. Rev. Lett. {\bf 87} (2001)
201801;
M. Cvetic, G. Shiu and A.M. Uranga, ``Chiral Four-Dimensional ${\cal N}=1$
Supersymmetric Type IIA Orientifolds from Intersecting D6 Branes,''
Nucl. Phys. {\bf B615} (2001) 3, hep-th/0107166; 
D. Berenstein, V. Jejjala and R. Leigh, ``The Standard Model on a D-Brane,''
Phys. Rev. Lett. {\bf 88} (2002) 071602, hep-th/0105042.}

\lref\Rabadan{L. Ibanez, R. Rabadan and A. Uranga, ``Anomalous $U(1)$'s in
Type I and Type IIB $D=4$, ${\cal N}=1$ String Vacua,'' Nucl. Phys.
{\bf B542} (1999) 112, hep-th/9808139.} 

\lref\oldgauge{M. Dine, W. Fischler and M. Srednicki, ``Supersymmetric
Technicolor,'' Nucl. Phys. {\bf B189} (1981) 575; S. Dimopoulos and
S. Raby, ``Supercolor,'' Nucl. Phys. {\bf B192} (1981) 353;
M. Dine and W. Fischler, ``A Phenomenological Model of Particle
Physics based on Supersymmetry,'' Phys. Lett. {\bf B110} (1982) 227;
C. Nappi and B. Ovrut, ``Supersymmetric Extension of the $SU(3) \times
SU(2) \times U(1)$ Model,'' Phys. Lett. {\bf B113} (1982) 175;
L. Alvarez-Gaume, M. Claudson and M. Wise, ``Low-Energy Supersymmetry,''
Nucl. Phys. {\bf B207} (1982) 96; S. Dimopoulos and S. Raby, ``Geometric
Hierarchy,'' Nucl. Phys. {\bf B219} (1983) 479.}

\lref\gepner{T. Dijkstra, L. Huiszoon and A. Schellekens, ``Chiral
Supersymmetric Standard Model Spectra from Orientifolds of Gepner Models,''
Phys. Lett. {\bf B609} (2005) 408, hep-th/0403196;
``Supersymmetric Standard Model Spectra from RCFT Orientifolds,''
Nucl. Phys. {\bf B710} (2005) 3, hep-th/0411129.}

\lref\sufive{I. Affleck, M. Dine and N. Seiberg, ``Dynamical Supersymmetry
Breaking in Chiral Theories,'' Phys. Lett. {\bf B137} (1984) 187.}

\lref\Brodie{J. Brodie, ``On Mediating Supersymmetry Breaking in D-Brane
Models,'' hep-th/0101115.}

\lref\soten{I. Affleck, M. Dine and N. Seiberg, ``Exponential Hierarchy
from Dynamical Supersymmetry Breaking,'' Phys. Lett. {\bf B140} (1984) 59.}

\lref\hitoshi{H. Murayama, ``Studying Noncalculable Models of Dynamical
Supersymmetry Breaking,'' Phys. Lett. {\bf B355} (1995) 187, hep-th/9505082.}

\lref\GSW{For a pedagogical discussion with references, see:
M. Green, J. Schwarz and E. Witten, {\it Superstring Theory:
Volume II}, Cambridge University Press, 1987.}

\lref\Ftheory{C. Vafa, ``Evidence for F-Theory,'' Nucl. Phys. {\bf 469}
(1996) 403, hep-th/9602022.}

\lref\Fmore{M. Bershadsky, K. Intriligator, S. Kachru, D. R. Morrison, V. Sadov and C. Vafa,
``Geometric Singularities and Enhanced Gauge Symmetries,'' Nucl. Phys. {\bf B481} (1996) 215,
hep-th/9605200.}

\lref\KKLT{S. Kachru, R. Kallosh, A. Linde and S. Trivedi, ``de Sitter
Vacua in String Theory,'' Phys. Rev. {\bf D68} (2003) 046005, hep-th/0301240.}

\lref\Sandip{J. Lykken, E. Poppitz and S. Trivedi, ``Branes with GUTS and
Supersymmetry Breaking,'' Nucl. Phys. {\bf B543} (1999) 105, hep-th/9806080.}

\lref\DDF{
F. Denef, M. R. Douglas and B. Florea, ``Building a Better Racetrack,''
JHEP {\bf 0406} (2004) 034, hep-pth/0404257; F. Denef, M. R. Douglas, B. Florea,
A. Grassi and S. Kachru, ``Fixing All Moduli in a Simple F-Theory
Compactification,'' hep-th/0503124;
D. L\"ust, S. Reffert, W. Schulgin and S. Stieberger, ``Moduli
Stabilization in Type IIB Orientifolds (I): Orbifold Limits,''
hep-th/0506090.
}

\lref\otherstab{V. Balasubramanian, P. Berglund, J. Conlon and F. Quevedo,
``Systematics of Moduli Stabilization in Calabi-Yau Flux Compactifications,''
JHEP {\bf 0503} (2005) 007, hep-th/0502058;
P. S. Aspinwall and R. Kallosh, ``Fixing All Moduli for M-Theory on K3 $\times$ K3, hep-th/0506014.}

\lref\GKLM{S. Gukov, S. Kachru, X. Liu and L. McAllister, ``Heterotic Moduli
Stabilization with Fractional Chern-Simons Invariants,'' Phys. Rev. {\bf D69} (2004)
086008, hep-th/0310159.}

\lref\othhet{G. Curio, A. Krause and D. L\"ust, ``Moduli Stabilization
in the Heterotic/IIB Discretuum,'' hep-th/0502168;
R. Brustein and S.P. de Alwis, ``Moduli Potentials in String Compactifications
with Fluxes: Mapping the Discretuum,'' Phys. Rev. {\bf D69} (2004) 126006,
hep-th/0402088.}

\lref\IIA{J. Derendinger, C. Kounnas, P. Petropoulos and F. Zwirner,
``Superpotentials in IIA Compactifications with General Fluxes,''
Nucl. Phys. {\bf B715} (2005) 211, hep-th/0411276;
S. Kachru and A.-K. Kashani-Poor, ``Moduli Potentials in Type IIA
Compactifications with RR and NS Flux,'' JHEP {\bf 0503} (2005) 066,
hep-th/0411279;
G. Villadoro and F. Zwirner, ``$\CN=1$ Effective Potential from Dual Type IIA
D6/O6 Orientifolds with General Fluxes,'' JHEP {\bf 0506} (2005) 047,
hep-th/0503169;
O. DeWolfe, A. Giryavets, S. Kachru and W. Taylor, ``Type IIA
Moduli Stabilization,'' JHEP {\bf 0507} (2005) 066, hep-th/0505160;
T. House and E. Palti, ``Effective Action of (Massive) IIA on Manifolds
with $SU(3)$ Structure,'' Phys. Rev. {\bf D72} (2005) 026004,
hep-th/0505177;
P.G. Camara, A. Font and L.E. Ibanez, ``Fluxes, Moduli Fixing and MSSM-like
Vacua in a Simple IIA Orientifold,'' JHEP {\bf 0509} (2005) 013, hep-th/0506066.}

\lref\oldbreak{V. Kaplunovsky and J. Louis, ``Model Independent Analysis of
Soft Terms in Effective Supergravity and String Theory,'' Phys. Lett.
{\bf B306} (1993) 269, hep-th/9303040;
A. Brignole, L. Ibanez and C. Munoz, ``Towards a Theory of Soft Terms
for the Supersymmetric Standard Model,'' Nucl. Phys. {\bf B422} (1994) 125,
hep-ph/9308271.}

\lref\FMW{R. Friedman, J. Morgan and E. Witten, ``Vector Bundles
and F-Theory,'' Comm. Math. Phys. {\bf 187} (1997) 679, hep-th/9701162.}

\lref\FMWII{R. Friedman, J. Morgan and E. Witten, ``Vector Bundles over Elliptic
Fibrations,'' alg-geom/9709029.}

\lref\OvrutII{V. Braun, Y.-H. He, B. Ovrut and T. Pantev, ``A Heterotic
Standard Model,'' Phys. Lett. {\bf B618} (2005) 252, hep-th/0501070;
``A Standard Model from the $E_8 \times E_8$ Heterotic Superstring,''
JHEP {\bf 0506} (2005) 039, hep-th/0502155.}

\lref\NDA{M. Luty, ``Naive Dimensional Analysis and Supersymmetry,''
Phys. Rev. {\bf D57} (1998) 1531, hep-ph/9706235; A. Cohen, D. B. Kaplan
and A. Nelson, ``Counting 4 pi's in Strongly Coupled Supersymmetry,''
Phys. Lett. {\bf B412} (1997) 301, hep-ph/9706275.}

\lref\SeibergNelson{A. Nelson and N. Seiberg, ``R-Symmetry Breaking
versus Supersymmetry Breaking,'' Nucl. Phys. {\bf B416} (1994) 46,
hep-ph/9309299.}

\lref\DineNelson{M. Dine, A. Nelson and Y. Shirman, ``Low-Energy
Dynamical Supersymmetry Breaking Simplified,'' Phys. Rev. {\bf D51} (1995)
1362, hep-ph/9408384;
M. Dine, A. Nelson, Y. Nir and Y. Shirman, ``New Tools for Low-Energy
Dynamical Supersymmetry Breaking,'' Phys. Rev. {\bf D53} (1996) 2658,
hep-ph/9507378.}

\lref\directmed{
K. Izawa, Y. Nomura, K. Tobe and T. Yanagida, ``Direct Transmission Models
of Dynamical Supersymmetry Breaking,'' Phys. Rev. {\bf D56} (1997) 2886,
hep-ph/9705228;
H. Murayama, ``A Model of Direct Gauge Mediation,''
Phys. Rev. Lett. {\bf 79} (1997) 18, hep-ph/9705271. }

\lref\BLT{R. Blumenhagen, D. L\"ust and  T. R. Taylor,
``Moduli Stabilization in Chiral Type IIB Orientifold Models with Fluxes,''
Nucl. Phys. {\bf B663} (2003) 319, hep-th/0303016.}

\lref\CU{J. F. G. Cascales and A. Uranga,
``Chiral $4D$ String Vacua with D-Branes and NSNS and RR Fluxes,''
JHEP {\bf 0305} (2003) 011, hep-th/0303024.}

\lref\LP{M. Larosa and G. Pradisi, ``Magnetized Four-Dimensional $\IZ_2 \times \IZ_2$ Orientifolds,''
Nucl. Phys. {\bf B667} (2003) 261, hep-th/0305224.}

\lref\FI{A. Font and L. Ibanez, ``SUSY Breaking Soft Terms in a MSSM Magnetized D7-Brane Model,''
JHEP {\bf 0503} (2005) 040, hep-th/0412150.}

\lref\MS{F. Marchesano and G. Shiu, ``MSSM Vacua from Flux Compactifications,''
Phys.Rev. {\bf D71} (2005) 011701, hep-th/0408059; ``Building MSSM Flux Vacua,''
JHEP {\bf 0411} (2004) 041, hep-th/0409132.}

\lref\CL{M. Cvetic and T. Liu, ``Supersymmetric Standard Models, Flux Compactification and Moduli Stabilization,''
Phys. Lett. {\bf B610} (2005) 122, hep-th/0409032;
M. Cvetic, T. Li and T. Liu, ``Standard-like Models as Type IIB Flux Vacua,''
Phys. Rev. {\bf D71} (2005) 106008, hep-th0501041.} 

\lref\DGS{D.-E. Diaconescu, A. Garcia-Raboso and K. Sinha, work in progress.}

\lref\SenI{A. Sen, ``F-Theory and Orientifolds,'' Nucl. Phys. {\bf B475} (1996) 562,
hep-th/9605150.}
\lref\SenII{A. Sen, ``Orientifold Limit of F-Theory Vacua'', Phys. Rev. {\bf D55}
(1997) 7345, hep-th/9702165.}
\lref\CurD{G. Curio and R. Donagi, ``Moduli in $\CN=1$ Heterotic/F-Theory Duality,''
Nucl. Phys. {\bf B518} (1998) 603, hep-th/9801057.}
\lref\Curio{G. Curio, ``Chiral Matter and Transitions in Heterotic String Models,''
Phys. Lett. {\bf B435} (1998) 39, hep-th/9803224.}
\lref\DLOPI{R. Donagi, A. Lukas, B. Ovrut and D. Waldram,
''Nonperturbative Vacua and Particle Physics in M-Theory,'' JHEP {\bf 9905} (1999) 018,
hep-th/9811168.}
\lref\DLOPII{R. Donagi, A. Lukas, B. Ovrut and D. Waldram,
``Holomorphic Vector Bundles and Nonperturbative Vacua in M-Theory,''
JHEP {\bf 9906} (1999) 034, hep-th/9901009.}
\lref\DOPWI{R. Donagi, B. Ovrut, T. Pantev and D. Waldram,
``Standard Models from Heterotic M-theory,'' Adv. Theor. Math. Phys. {\bf 5} (2002) 93,
hep-th/9912208.}
\lref\DOPWII{R. Donagi, B. Ovrut, T. Pantev and D. Waldram,
``Standard Model Bundles,'' Adv. Theor. Math. Phys. {\bf 5} (2002) 563, math.AG/0008010.}
\lref\OPP{ B. Ovrut, T. Pantev and J. Park, ``Small Instanton Transitions in Heterotic M-Theory,''
JHEP {\bf 0005} (2000) 045, hep-th/0001133.}
\lref\BDO{E. I. Buchbinder, R. Donagi and B. Ovrut, ``Vector Bundle Moduli and Small Instanton Transitions,''
JHEP {\bf 0206} (2002) 054, hep-th/0202084.}
\lref\DR{D.-E. Diaconescu and G. Rajesh,
``Geometrical Aspects of Fivebranes in Heterotic/F-Theory Duality in Four Dimensions,''
JHEP {\bf 9906} (1999) 002, hep-th/9903104.}
\lref\MVII{D. R. Morrison and C. Vafa, ``Compactifications of F-Theory on Calabi--Yau Threefolds -- II,''
Nucl. Phys. {\bf B476} (1996) 437, hep-th/9603161.}
\lref\ATasi{P. S. Aspinwall, ``K3 Surfaces and String Duality,'' hep-th/9611137.}
\lref\Apoint{P. S. Aspinwall and M. Gross, ``The $SO(32)$ Heterotic String on a K3 Surface,''
Phys.Lett. {\bf B387} (1996) 735, hep-th/9605131; P. S. Aspinwall, ``Point-like Instantons and the 
$Spin(32)/\IZ_2$ Heterotic String,'' Nucl. Phys. {\bf B496} (1997) 149, hep-th/9612108.}
\lref\AM{P. S. Aspinwall and D. R. Morrison, ``Point-like Instantons on K3 Orbifolds,''
Nucl. Phys. {\bf B503} (1997) 533, hep-th/9705104.}
\lref\JK{J. Kollar, ``The Structure of Algebraic Threefolds: an Introduction to Mori's Program,''
Bull. Amer. Math. Soc. {\bf 17} (1987) 211.}
\lref\DHOR{R. Donagi, Y.-H. He, B. Ovrut and R. Reinbacher,
``The Particle Spectrum of Heterotic Compactifications,''
JHEP {\bf 0412} (2004) 054, hep-th/0405014.}
\lref\BJPS{M. Bershadsky, A. Johansen, T. Pantev and V. Sadov,
``On Four-Dimensional Compactifications of F-Theory,'' Nucl. Phys. {\bf B505} (1997) 165,
 hep-th/9701165.}
\lref\DI{D.-E. Diaconescu and G. Ionesei, `` Spectral Covers, Charged Matter and Bundle Cohomology,''
JHEP {\bf 9812} (1998) 001, hep-th/9811129.}
\lref\G{A. Grassi, ``Divisors on Elliptic Calabi-Yau 4-folds and the Superpotential in F-Theory -- I,''
alg-geom/9704008.}
\lref\KN{B. V. Karpov and D. Yu. Nogin, ``Three-block Exceptional Collections over del Pezzo Surfaces,''
alg-geom/9703027.}
\lref\AC{B. Andreas and G. Curio, ``On Discrete Twist and Four-Flux in $\CN=1$ Heterotic/F-Theory Compactifications,''
Adv. Theor. Math. Phys.  {\bf 3} (1999) 1325, hep-th/9908193.}
\lref\DFR{M. R. Douglas, B. Fiol and  C. R\"omelsberger, ``Stability and BPS Branes,''
JHEP {\bf 0509} (2005) 006, hep-th/0002037.}
\lref\Dcat{M. R. Douglas, ``D-Branes, Categories and $\CN=1$ Supersymmetry,'' J. Math. Phys. {\bf 42} (2001) 2818,
hep-th/0011017.}
\lref\AD{P. S. Aspinwall and M. R. Douglas, ``D-Brane Stability and Monodromy,''
JHEP {\bf 0205} (2002) 031, hep-th/0110071.}
\lref\Di{R. Donagi, ``Principal Bundles on Elliptic Fibrations,'' alg-geom/9702002.}
\lref\Aiii{P. S. Aspinwall, ``Aspects of the Hypermultiplet Moduli Space in String Duality,'' JHEP {\bf 9804} (1998) 019,
hep-th/9802194.}
\lref\ACiii{B. Andreas and G. Curio, ``Horizontal and Vertical Fivebranes in Heterotic/F-Theory Duality,'' JHEP {\bf 0001}
(2000) 013, hep-th/9912025.}
\lref\Miranda{R. Miranda, ``Smooth Models for Elliptic Threefolds'', in {\it The Birational Geometry of
Degenerations,} R. Friedman and D.R. Morrison eds., Progress in Mathematics {\bf 29}, Birkh\"auser
1983.}
\lref\CSi{P. Candelas and H. Skarke, ``F-Theory, $SO(32)$ and Toric Geometry,'' Phys. Lett. {\bf B413} (1997) 63, 
hep-th/9706226.}

\Title{
\vbox{
\baselineskip12pt
\hbox{hep-th/0512170}
\hbox{SLAC-PUB-11593}
\hbox{SU-ITP-05/33}}}
{\vbox{\vskip 2pt
\vbox{\centerline{Gauge-Mediated Supersymmetry Breaking}}
\vskip2.5pt
\vbox{\centerline{in String Compactifications}}
}}
\vskip 11pt
\centerline{Duiliu-Emanuel Diaconescu\footnote{$^\forall$}{{{\tt duiliu@physics.rutgers.edu}}}$^1$,
Bogdan Florea\footnote{$^{\sharp}$}{{{\tt bflorea@slac.stanford.edu}}}$^2$, Shamit Kachru\footnote
{${^\natural}$}{{{\tt skachru@stanford.edu}}}$^2$ and
Peter Svr\v{c}ek\footnote{${^\flat}$}{{{\tt svrcek@slac.stanford.edu}}}$^2$}
\bigskip
\medskip
\centerline{{\it $^1$Department of Physics and Astronomy, Rutgers University,}}
\centerline{{\it Piscataway, NJ 08855-0849, USA}}
\centerline{{\it $^2$Department of Physics and SLAC, Stanford University,}}
\centerline{\it Stanford, CA 94305/94309, USA}
\bigskip
\bigskip
\bigskip
\noindent
We provide string theory examples where a toy model of a SUSY GUT or the MSSM
is embedded in a compactification along with a gauge sector which dynamically
breaks supersymmetry.  We argue that by changing microscopic details of the model
(such as precise choices of flux), one can arrange for the dominant mediation mechanism
transmitting SUSY breaking to the Standard Model to be either gravity mediation or gauge
mediation. Systematic improvement of such examples may lead to top-down models
incorporating a solution to the SUSY flavor problem.

\vfill
\Date{December 2005}

\newsec{Introduction}

The Minimal Supersymmetric Standard Model (MSSM), and its natural extension
into SUSY GUTs, provides perhaps the most compelling viable extension of the
Standard Model \refs{\MSSM,\Nilles}.  Supersymmetric models
can stabilize the hierarchy between
the electroweak and Planck scales, and also successfully incorporate
gauge coupling unification \unification.
One can further hope that the correct theory of supersymmetry breaking
explains the small scale of breaking via a dynamical mechanism \WittenDSB,
so that the weak scale is not only radiatively stable, but is also
explained in a theory with no (very) small dimensionless parameters.

From the top down, supersymmetric GUTs have seemed very natural in the
context of heterotic string theory \GSW.
It has been known for some time that explicit models with
pseudo-realistic matter content can be constructed in this framework;
the state of the art models are presented in e.g. \OvrutII.
In the heterotic M-theory framework \HW, one can also accommodate
gauge-gravity unification in a fairly natural way \WittGG.

One missing ingredient in many of the stringy constructions has been
an explicit model of supersymmetry breaking.
In the old heterotic framework, one fruitful approach was to simply
parameterize the SUSY breaking by assuming (without microphysical
justification) that the dominant SUSY-breaking
F-term arises in the dilaton or a given modulus multiplet.  Then, using
the structure of the low-energy supergravity, one can work out the patterns
of soft terms in different scenarios \oldbreak.
More recently, type II flux vacua with intersecting D-branes
have become a popular arena for
phenomenological constructions as well \Dreview. In these models, the fluxes
generate calculable SUSY breaking F-terms in various circumstances
\refs{\softsusy,\FI,\MS,\CL}.  A
full model of the soft terms must necessarily also
solve the problem of moduli stabilization;
by now there exist type IIB and type IIA constructions where this
problem is solvable \refs{\KKLT,\DDF,\EvaAlex,\IIA,\otherstab}.  In all
such constructions, of course,
one must fine-tune the cosmological constant after SUSY breaking.  This
tune has always been performed in the phenomenological
literature by a shift of the
constant term in the superpotential $W$, and
the flux discretuum \BP\ seems to microphysically
allow the same procedure in string theory.\foot{Since in realistic
models incorporating the MSSM or its extensions,
the primordial scale of SUSY breaking is always $\leq
10^{11}$ GeV, the required constant in $W$ is always parametrically small
relative to the Planck or string scale.  This can be viewed as a ``bottom up"
motivation for the small $W_0$ tune which is performed in many models of moduli
stabilization \KKLT, where the same tune allows one to stabilize moduli
in a calculable regime.}

One notable problem with the soft-terms induced by the fluxes is that their
natural order of magnitude is typically only suppressed from the string scale
by a few powers of the Calabi-Yau radius.
While this may lead to suitable soft-terms in models with significant warping
\refs{\Herman,\GKP} or with a low string scale, both of these solutions destroy
one of the main attractive features of supersymmetry -- its natural connection
to grand unified models.

In this paper, we describe some simple pseudo-realistic string constructions,
which incorporate both a toy model of the MSSM or a SUSY GUT,
and also incorporate a sector which accomplishes
dynamical SUSY breaking (for early attempts in this direction, see e.g.
\Sandip).
It will be clear
that our basic setup is sufficiently
modular that one can view the MSSM/GUT sector as a ``black box,'' and could
presumably improve the realism in that sector without disturbing the
basic mechanism of SUSY breaking or mediation.
We give two classes of constructions:
one based on the so-called ``non-calculable" SUSY breaking models of
\refs{\sufive,\soten}, and another
based on the recent insight that
there are plentiful examples of
simple quiver gauge theories which exhibit dynamical
supersymmetry breaking (DSB) \quiverDSB, and which
can be easily embedded in Calabi-Yau compactification.
By tuning closed-string parameters (in particular, choices of background
flux), we will argue that one can find in each case two different
regimes: one regime where the dominant mediation mechanism transmitting
SUSY breaking to the Standard Model is gravity mediation, and another
where it is gauge mediation.

In different parts of the paper, we will use the language and
techniques of heterotic string model building and of F-theory (or
IIB string theory) constructions.  A wide class of 4d
$\CN=1$ models admit dual heterotic and F-theory descriptions, and we
simply use whichever description is more convenient in a given
circumstance.  In many cases, one should be able to use the
dictionary of \FMW\ to translate back and forth. Because most of the
models of supersymmetry breaking in the string literature have
involved gravity mediation, in \S2 we briefly review some elementary
phenomenology, explaining the chief differences between gravity and
gauge mediation. In \S3, we recall basic features of the
``non-calculable'' models of DSB
\refs{\sufive,\soten} and give examples where these can be embedded
into string theory along with a toy-model of a SUSY GUT. In \S4, we
briefly review the construction of SUSY breaking quiver gauge
theories in non-compact Calabi-Yau manifolds \quiverDSB. We turn to
the embedding into a compact geometry containing both a toy-model
SUSY GUT and a SUSY breaking sector in \S5, while in \S6 we
construct compact models incorporating a SUSY breaking sector and
the MSSM-like theory of \Verlinde\ (presumably, similar constructions
could be given incorporating the semi-realistic models of
\refs{\bottomup,\gepner}
or other attempts at constructing the SM on intersecting branes or
branes at singularities).
In a concluding section, we describe several promising directions for further
research.

We should state clearly at the outset that our
``semi-realistic'' explicit constructions are
not close to being fully realistic.
However, it seems clear that systematic further work along
these lines could produce increasingly realistic models.
Similarly, the constructions even at this level of realism are rather
complicated, and many issues beyond those which we discuss
(related to both gauge theory model building and to moduli stabilization)
could be explored in each toy model.
We will explicitly point out our assumptions, and our
justification for making these assumptions, at various points in the
text.

\newsec{Gauge Mediation versus Gravity Mediation}

In gauge mediated models of SUSY breaking,
SUSY is broken in a hidden sector with gauge group $G_H$.
The SUSY breaking in the hidden sector leads to
splittings for a vector-like set of  messenger chiral multiplets
$\tilde \phi_i, \phi_i$, which carry Standard
Model gauge charges.  Standard Model gauge interactions then lead to
a one-loop gaugino mass and a two-loop mass$^2$ for the other sparticles.
In
many models, the $\phi_i$ are neutral under $G_H$
but are coupled to the SUSY breaking by additional gauge singlets; in
other
models of ``direct mediation,'' the $\phi_i$ can be charged under $G_H$.
Some classic references include \refs{\oldgauge,\DineNelson,\GR}.

The primary virtue of models of gauge mediation is that they solve the
SUSY flavor problem, the problem of why the
soft-breaking terms do not introduce
new sources of flavor violation which violate present experimental bounds.
The fact that gauge mediation generates e.g. universal squark masses is
clear, because the only coupling of the squarks
to the messengers occurs through
(universal) gauge interactions.  While a similar slogan might naively be
applied to gravity mediation, in fact it has been understood
that generic gravity mediated models do ${\it not}$ enjoy such universality.
Suppose, in such a scenario, $X$ is the
modulus whose F-term breaks SUSY, and $Q_i$ denote generic SM fields with
$i$ running over generations.  Then, couplings of the
form
\eqn\dimsix{K = \int d^4 \theta {c_i\over M_P^2} X^\dagger X Q_i^{\dagger} Q_i
+ \cdots}
in the K\"ahler potential
exist in generic models.  In string constructions, they arise by integrating out
massive fields to write down the 4d effective Lagrangian.  These operators
occur with different ${\cal O}$(1) coefficients $c_i$, and generally yield
non-universal squark masses.\foot{Exceptions exist: for instance, dilaton
domination in the weakly coupled heterotic string, can yield universal soft
masses \oldbreak.}
From an effective field theory perspective, this failure of universality is
easily understood \HKR: gravity mediation is sensitive to Planck-scale physics,
and the physics of flavor (which is presumably determined by the geometry
of the compactification manifold) is visible to the massive fields which
are integrated out to yield \dimsix.

In gauge mediated models, on the other hand, one begins with an effective
superpotential
\eqn\gaugemed{W = \tilde \phi_i X \phi_i + W_{MSSM}}
where $X$ is a spurion superfield whose
vev
\eqn\xvev{X = M + \theta^2 F}
both gives the messengers a mass, and breaks supersymmetry.  One can think
of \gaugemed\ as an effective theory that parameterizes the piece of hidden
sector physics relevant to the Standard Model; only the $\phi_i$
are charged under $SU(3) \times SU(2) \times U(1)$, so one can ``integrate out''
the hidden sector, parameterizing its effects via \gaugemed, \xvev.
A standard analysis \GR\ then yields the sparticle masses and $A$-terms
in terms of $F/M$ and the ``messenger index'' $N$ (basically, the number of
messenger fields).
Very roughly, one finds squark and slepton masses $m_Q^2 \sim
\alpha^2 F^2/M^2$ and
gaugino masses $m_{\lambda} \sim \alpha F/M$;
we shall describe the results in more detail for our particular toy models
in later sections.

The messenger fields are charged under the Standard Model, and contribute
to running of the gauge couplings above their mass $M$.  Hence,
in gauge mediated
models, one finds a shifted value of the unified gauge coupling:
\eqn\gutch{\delta \alpha_{GUT}^{-1} = -{N\over 2\pi} {\rm ln} {M_{GUT}\over M}}
where
\eqn\nis{N = \sum_{i} n_i}
and $n_i$ is twice the Dynkin index of the observable sector
gauge representation ${\bf r}_i$ of
the $i$th messenger.
One can determine the maximal value of $N$ (consistent with weakly coupled
unification) as follows.

If one wants $\alpha F/M \sim {\rm TeV}$, it follows one should take
$F/M \sim 10-100$ TeV.  Then, for the highest F-term consistent with dominance
of gauge mediation over gravity mediation $F \sim (10^{10} {\rm GeV})^2$, one
would have $M \sim 10^{15}$ GeV.  Perturbativity of gauge interactions requires
\eqn\bound{N \leq 150/ {\rm ln} {M_{GUT}\over M} ~.}
Hence, as the messenger scale and SUSY-breaking F-term increase, one is allowed
a rather large number of messengers.  However, phenomenological considerations
(such as the desire to avoid a gravitino problem, in addition to the need to
keep the relative significance of
the gravity mediated contribution sufficiently small)
generally favor $F$ terms below $10^{10}$ GeV, and
a correspondingly smaller number
of messengers.  For $M=10^{10}$ GeV one obtains a bound on
$N$ of about 10 (which would correspond to five ${\bf{5} + \overline{\bf{5}}}$ pairs
if the SM is embedded in an $SU(5)$ GUT; for purposes of discussion we will
always assume this) \GR.

\subsec{Strategy for embedding into string theory}

Perhaps the most obvious place to try and construct a GUT model with
supersymmetry breaking would be the strongly coupled heterotic
string.  Indeed, some of our toy models will have an explicit
heterotic realization.  However, we will also provide a dual
F-theory description; in the F-theory formalism, the physics which
controls moduli stabilization is better understood, since simple
ingredients like NS and RR fluxes which generate potentials in type
II strings, dualize to rather intractable ingredients in the
heterotic theory.

It is well known that heterotic string compactifications on elliptically
fibered Calabi-Yau $n$-folds, are dual to F-theory (type IIB) models on
K3-fibered Calabi-Yau $(n+1)$-folds.  For $n=1$, this was described in
\Ftheory, while a detailed map for $n=2$ was provided in \Fmore.
For $n=3$ the story is considerably more involved, but some important
aspects involving nontrivial gauge bundles (relevant for constructing
GUT models) were worked out in a series
of papers by Friedman, Morgan and Witten \FMW.
It is important to note that many of the
pseudo-realistic heterotic models constructed by
the Penn group (see \DHOR\ and refs therein) involve the spectral cover
construction on elliptic threefolds, and hence fall squarely into the class of
constructions which admit dual F-theory descriptions.
Because it will be easier to use the F-theory description to also make a model
of gauge mediation in the later
sections of this paper, we will mostly stick to the D-brane/F-theory language.

So, we will try to engineer a hidden SUSY breaking sector
on a stack of D-branes in F-theory.
The grand unified extension of the Standard Model or brane MSSM, will live in
different constructions on either
a stack of D7-branes (dual to one of the heterotic $E_8$ walls,
in some cases), or D3-branes at a singular point in the Calabi-Yau space.
SUSY will be broken non-perturbatively at the dynamical scale of the hidden
sector gauge theory,
$\Lambda_H$.
This can quite naturally be a scale which is parametrically low compared to
$M_P$.  The detailed geometries which accomplish this are discussed in
later sections.

\medskip
\noindent{\it{Transmission to the Standard Model}}
\medskip

The dominant interactions which transmit SUSY breaking to the
observable sector, will depend on the distance between the hidden
and observable brane stacks. In the F-theory construction, there are
complex structure moduli of the Calabi-Yau fourfold which control
the brane positions.  These can be interpreted as e.g. singlets
or adjoints in
the D7-brane gauge theory.\foot{In perturbative heterotic string GUTs,
adjoint representations cannot appear in gauge groups arising from
level 1 worldsheet current algebras.  However in F-theory constructions, 
D3 and D7 brane gauge theories can often contain adjoint matter fields.
These generally dualize to non-perturbative sectors in the heterotic
theory.  This raises the possibility of constructing D7 GUTs in F-theory
with a more conventional GUT-breaking mechanism replacing symmetry
breaking by Wilson lines.}
These
singlets or adjoints are generically stabilized (i.e. given a mass) by
background flux in the F-theory description \GKTT. Depending on
where in their moduli space the D7 adjoints are stabilized, the
distance $d$ between the SUSY breaking and observable sector brane
stacks will vary. For flux choices which stabilize these stacks
(i.e. the fourfold complex structure) in a regime where
\eqn\smalld{{d\over \alpha^\prime} = M << M_s} the dominant
interaction between the brane stacks is via zero modes of open
strings stretching between them. In four dimensions, the open
strings are described by chiral fields $\phi_i,\tilde\phi_i$
transforming in conjugate representations of the GUT gauge group.
For example, if we assume an $SU(5)$ GUT coming from a stack of five
D7-branes, $\phi_i,\tilde\phi_i$ transform in ${\bf 5}\oplus{\bf \bar
5}$ representation of $SU(5).$ The number of copies of the
representations depends on the details of the hidden gauge group
supported on the other D7-brane stack. In the effective four
dimensional language, these fields are messengers of gauge mediated
supersymmetry breaking. Hence, one obtains gauge mediation with
messenger scale $M$ equal to the mass of the stretched strings.

To confirm the reasonableness of assuming flux stabilization in the
parameter regime \smalld, one could do a simple statistical analysis
along the lines of \Douglas. Existing results about similar mild
tunes make it fairly clear that the regime \smalld\ should be
attainable for the phenomenologically relevant range of values of
$M$, but a more detailed statistical analysis might be interesting.

If the D-branes have separation $d$ such that \eqn\grseparation{
{d\over \alpha'}>>M_s,} supersymmetry breaking is mainly
communicated via zero modes of the closed strings. This corresponds
to gravity or moduli mediation.  For ${d/\alpha'}\sim M_s,$  the
D-branes stacks have string scale separation. The four dimensional
effective description breaks down since the D-brane stacks interact
via the whole tower of excited string oscillators. The supersymmetry
breaking from the hidden sector to our braneworld is ``string
mediated.'' In this way, string theory unifies different mechanisms
of supersymmetry breaking in a single string compactification. The
mediation is primarily via gauge or gravitational interactions,
depending on the
distance between the brane stacks.

\newsec{Noncalculable Models of Supersymmetry Breaking}

The standard lore about SUSY breaking in the hidden sector of the heterotic
string, involves the assumption that hidden $E_8$ gaugino condensation can be
responsible for SUSY breaking.  (More complicated models with ``racetrack''
potentials are also commonly discussed).  However, as argued convincingly
in \SeibergNelson, and as is clear from analysis of the relevant effective
potentials after the tree-level no-scale structure is broken (see e.g.
\refs{\GKLM,\othhet}), generically hidden $E_8$
gaugino condensation does not guarantee supersymmetry breaking.
This is not terribly surprising, since in the flat space limit gaugino
condensation in $\CN=1$ field theory does not break supersymmetry.  Here,
and in subsequent sections,
we describe some examples where the hidden sector breaks supersymmetry even
in the flat space limit, and the embedding into string theory
will not (in any model where closed-string moduli are stabilized)
relax the SUSY breaking F-term.\foot{In the
paper \KKLT, an anti-D3 brane in a warped background \KPV\ is used
to induce an exponentially small scale of supersymmetry breaking.
While
for many purposes such models may be quite similar to models incorporating
dynamical supersymmetry breaking \Dine, the field theoretic description of the
present class of models is certainly more transparent.  Of course, many
other ways of accomplishing supersymmetry breaking and yielding a positive
contribution to the potential in the presence of moduli stabilization
have also been studied in detail by now.}

\subsec{Simple alternatives to the hidden $E_8$}

We would like to choose a simple mechanism of SUSY breaking that is
easily embedded into string theory.  One of the lessons of
\refs{\KKLT,\DDF} is that it is possible to fix the geometrical
moduli of F-theory compactifications ${\it supersymmetrically}$. (As
mentioned previously, the small $W_0$ assumed there can perhaps be
thought of as the small constant $W$ that will be needed after SUSY
breaking, to cancel the cosmological constant). We will therefore
assume that all Calabi-Yau moduli are fixed as in those papers with
masses close to the string scale (before any further ingredients
which can yield dS vacua are considered), and
look for our SUSY breaking F-term
elsewhere.\foot{More precisely, the complex and dilaton moduli have masses
which scale as ${{\alpha^\prime} \over R^3}$ where $R$ is the
Calabi-Yau radius; the \kah\ moduli may have
masses which are significantly smaller and $W_0$ dependent.}

Studies of dynamical SUSY breaking models in the mid 1980s yielded a
particularly simple class of models, often called ``non-calculable
models.''  The simplest examples are the $SU(5)$ gauge theory with one
generation of ${\bf \overline 5}\oplus {\bf 10},$ and the $SO(10)$
gauge theory with one generation of ${\bf 16}.$ These models were
found by the following logic. Consider an $\CN=1$ supersymmetric
gauge theory without flat directions at tree level, and with
sufficiently little matter content that it is expected to undergo
confinement. 't Hooft anomaly matching for all global symmetries of
the theory can then constrain the possible low-energy pion
Lagrangians which describe the theory in the IR. In some cases, the
possible spectra that saturate the anomalies (under the assumption
that the symmetries are unbroken) are so contrived-looking that it
is implausible that the theory generates such composites; in such a
case, one must postulate that the global symmetries are broken in
the IR.  This means that there must be Goldstone bosons.  But
unbroken supersymmetry would require that they be complexified into
full chiral multiplets whose scalar vevs are unconstrained.  The
existence of suitable partners to complexify the Goldstones is very
implausible in a theory without tree-level flat directions.  Hence,
the theory must spontaneously break supersymmetry.

Following this logic, both the $SO(10)$ gauge theory with a single ${\bf 16}$
\soten,
and the $SU(5)$ gauge theory with a single ${\bf {\overline 5}} \oplus {\bf 10}$
\sufive, should be expected to exhibit dynamical supersymmetry breaking.
Further evidence that these theories do indeed dynamically break supersymmetry
was provided in \hitoshi\ by adding vector-like matter multiplets and computing
the Witten index and vacuum structure.
By now the case that SUSY breaking indeed occurs is quite compelling.
Many other such models exist, but these are the simplest cases and we will
be satisfied to use them for our toy constructions.

\medskip
\noindent{\it{Comments on the possibility of additional matter}}
\medskip
We should point out here that in general, geometric engineering of
the non-calculable models could also yield additional nonchiral
flavors in the ${\bf 5} \oplus  {\bf{\overline 5}}$ of $SU(5)$, or the
${\bf 10}$ of $SO(10)$.  Since these representations are nonchiral,
their appearance in the spectrum of fields with mass $< M_{string}$
will depend on the full details of moduli stabilization. We note
that even in the presence of such fields, the supersymmetry breaking
minimum persists \hitoshi. Because they are vector-like, one
generically expects worldsheet instantons (in the heterotic
description) and/or fluxes to lift their masses to a relatively high
scale.  However, their presence below the GUT scale can change the
RG running of the hidden sector gauge coupling, and hence the
scale at which supersymmetry breaks.

\subsec{Energy scales}

The supersymmetry in the chiral $SU(5)$ or $SO(10)$ gauge theories
is broken dynamically with F-term of the order of $F\sim (\Lambda_H/
4\pi)^2$, where $\Lambda_H$ is the strong coupling scale of the
hidden gauge theory \NDA.\

If the distance between the D7-brane stacks is much larger than the
string length $\ell_s,$ supersymmetry is gravity mediated. The
masses of the MSSM sparticles are of order \eqn\masem{m_s\sim
{F\over M_P}\sim {\Lambda_H^2\over (4\pi)^2 M_P}.} For $d<\ell_s$,
supersymmetry breaking is predominantly mediated by open strings
connecting the D-brane stacks. This is a stringy description of gauge
mediation, hence the sparticle masses are $m_s\sim \alpha (F/ M),$
where $M$ is the mass of the open strings acting as messengers of
supersymmetry breaking.

Hence, from the knowledge of gauge coupling $\alpha_H$ at the string
scale and the matter content of the hidden gauge group, we can
estimate the scale $\lambda_H$ at which the hidden gauge group gets
strongly coupled and breaks supersymmetry, and the masses of the
sparticles. At the string scale, the gauge couplings of the GUT and
of the hidden gauge group are approximately equal. One sees this
most easily in the heterotic description of the models, where this
comes from the equality of the gauge couplings of the two $E_8$'s of
the $E_8\times E_8$ heterotic string theory. In the following
discussion, we assume $\alpha_H=\alpha_{GUT}\approx 1/25$ at the
string scale $M_s=M_{GUT}=2\times 10^{16} ~\GeV$.  At the end we
discuss to what extent this approximation is valid in the F-theory
compactification constructed in the following section.

The one-loop RG evolution of $\alpha=g^2/4\pi$ is governed by
\eqn\rggovern{\mu{d \alpha^{-1}\over d\mu}={b\over 2\pi},} where $b$
in a supersymmetric gauge theory is $b=3C_2(G)-C(R).$ Evaluating
this for $SU(5)$ with one generation of ${\bf \overline 5}\oplus{\bf
10}$ gives\foot{Here we use that $C_2(SU(5))=5, C_2({\bf \overline
5})=1/2$ and $C_2({\bf 10})=3/2.$} $b=13.$  Hence the gauge theory
becomes strongly coupled at the scale
\eqn\scosca{\Lambda_{H}=M_{GUT}~ \exp(-2\pi/ 13\alpha_{GUT})\approx
10^{11}~ \GeV.} We estimate the F-term using naive dimensional
analysis \NDA\ to be $F\approx ({\Lambda_H\over 4\pi})^2\approx
\left(10^{10} ~\GeV\right)^2.$

For gauge mediation $F\sim (10^{10} ~\GeV)^2$ is at the upper end of
the range values for the F-term for which gauge mediation dominates
gravity mediation. This leads to sparticle masses of $\sim{\rm TeV}$
if the mass of the messenger particles is $M\approx 10^{15} ~\GeV.$
This is the largest $M$ for which the flavor problem can plausibly be
solved by gauge mediation (see eqn. (2.44) of \GR).
The messengers are open strings stretched between the two stacks of
D7-branes. If the branes are separated by distance $l,$ the
messenger mass is $M=2\pi l/g_s\ell_s^2.$ Assuming
$\ell_s=M_s^{-1}\approx2\times 10^{16} ~\GeV,$  the messenger mass
of $M=10^{15} ~\GeV$ corresponds to interbrane separation of
$l=10^{-2}g_s \ell_s.$\foot{We have neglected the contribution
of the messengers to the running of the hidden sector gauge coupling in
computing $\Lambda_H$,
since they decouple almost immediately given the high value of $M$.
Including their effects would lower $\Lambda_H$ slightly.}

In gravity mediation, this would lead to sparticle masses of the
order of $m_s\sim {F\over M_P}\sim 10^{-1} ~{\rm TeV}$ or less, which is a
bit low (as it should be, for the gauge mediated contribution to dominate the
mass squared matrix for the squarks). For gravity mediation, we
would prefer $F\sim (10^{11}
~\GeV)^2.$ This actually might be the case because the hidden gauge
group might be more strongly coupled than the GUT gauge group at the
string scale, as we discuss later.

As we said, $M \sim 10^{15}$ GeV is at the high end of the allowed
range for gauge mediation.
A scenario which would lead to lower $M$ with essentially the same
physics, is to imagine that there are some additional ${\bf 5} \oplus
{\bf{\overline 5}}$ pairs which are present in the sigma model tree-level
spectrum and receive a mass only from worldsheet instantons.  Such
extra vector-like states are quite common in heterotic constructions;
their presence or absence depends on the full details of the point in
moduli space chosen to study a given model.
The presence of $n$ extra pairs would reduce $b$ to $13-n$ for some
portion of the RG running starting from the GUT scale, and hence decrease
$\Lambda_H$.

The $SO(10)$ gauge theory with one chiral matter multiplet in the
${\bf 16}$ has\foot{This follows from the $SO(10)$ values of
$C_2(SO(10))=8,$ $C_2({\bf 16})=2.$} $b=22$, so the gauge theory
gets strongly coupled at a higher scale
$\Lambda_{SO(10)}=M_{GUT}~\exp(-2\pi/22\alpha_{GUT})\approx 10^{13}
~\GeV.$ The masses of the sparticles due to gravity mediation of
supersymmetry alone are $\sim10^{3} ~{\rm TeV}$ which is rather
high. The additional contribution to sparticle masses from gauge
mediation would make the masses even larger.
To make even a realistic model of gravity mediation with this hidden
sector, one would need to invoke the assumption of some extra ${\bf {10}}$'s of
$SO(10)$ with intermediate scale masses, to lower $\Lambda_H$.

In the above discussion, we neglected that the hidden gauge group
may actually be more strongly coupled than the GUT gauge group at
the string scale. In the heterotic description, which at finite
string coupling can be viewed as a compactification of M-theory on a
Calabi-Yau manifold times an interval, this is a consequence of
embedding the GUT symmetry into the larger of the two end of the
world-branes. The other world-brane has smaller volume because of
the warping along the interval. But the gauge couplings are
inversely proportional to the volumes of the world-branes
$\alpha={\ell_{11}^6/ V},$  where
$\ell_{11}=(4\pi\kappa_{11}^2)^{1/9}$ is the eleven-dimensional
Planck length. Hence, the hidden gauge group is more strongly
coupled at the string scale by a factor of $V_{GUT}/V_H$.

In the F-theory picture, the gauge symmetry comes from two stacks of
D7-branes.  The inverse gauge couplings of the observable and hidden sectors,
are controlled by the volumes of the two divisors that the stacks wrap,
which are two different sections of the $\IP^1$ fibration of the base of the
Calabi-Yau fourfold. The homology classes of the divisors cut out by
the two sections differ by the class of a $\IP^1$
fibration over some curve $\eta$ in the base. Hence, the volumes of the
two sections differ by an amount that grows with the size of the
$\IP^1$ fibre. In our example, the GUT symmetry comes from the divisor with
larger volume.

One of the implications of this is that we can increase the strength
of supersymmetry breaking effects by stabilizing the $\IP^1$ fibres at
larger volume. With larger $\IP^1$ fibres, the hidden gauge group
comes from a divisor with a smaller volume compared to the volume of
the GUT divisor, hence it is more strongly coupled and dynamically
breaks supersymmetry at a higher scale. In the one generation
$SU(5)$ model, this effect could increase the supersymmetry breaking
F-terms to the intermediate scale $10^{11} ~{\GeV}$ preferred in
the gravity mediated SUSY breaking solution to the hierarchy problem.
The $SO(10)$ model of a hidden sector leads to a somewhat high scale of
SUSY breaking if $\alpha_H=\alpha_{GUT}$ at the string scale.
Increasing the volume of the $\IP^1$ would only exacerbate this problem.

For gauge mediation, both the gravitino problem and the dominance
over gravity mediation prefer a lower scale of supersymmetry
breaking with $F\sim (10^{10}\GeV)^2$ being at the upper end of the
allowed values.
So while these models provide simple toy models where one can
compare the strength of gauge and gravity mediation, with gauge mediation
marginally winning in one case (and winning clearly if there are extra
vector-like representations of the hidden sector gauge group
at intermediate masses),
a clear next step would be to identify analogous hidden sectors which are
easy to engineer and give rise to much smaller $\Lambda_H$.

\subsec{Some stringy embeddings}

Following the above discussion, we will construct heterotic models with a three
generation $SU(5)$ GUT sector and a one generation $SU(5)$ hidden sector.

Consider heterotic $E_8\times E_8$ compactifications on smooth Weierstrass models
$\pi:Y\to S$. The base $S$ is a del Pezzo surface.
We would like to construct a background bundle $V_1\times V_2$ on $Y$ so that
both $V_1$ and $V_2$ are stable $SU(5)$ bundles and the following
conditions are satisfied

$I)$ $V_1, V_2$ yield one and respectively three generations i.e.
\eqn\generations{
\hbox{ch}_3(V_1)=\pm 1,\qquad \hbox{ch}_3(V_2)=\pm 3.}

$II)$ Anomaly cancellation:
\eqn\anomaly{
c_2(V_1)+c_2(V_2)+ \Lambda = c_2(Y)}
where $\Lambda$ is an effective curve on $Y$ which supports
background heterotic fivebranes \refs{\DLOPI,\DOPWI,\DLOPII}.
Motivated by K\"ahler moduli stabilization, we would like to
impose an additional condition on the fivebrane class $\Lambda$. Suppose $\Lambda$ has a decomposition
$$
\Lambda = \Xi + N_5E
$$
where $\Xi$ is a horizontal curve on $Y$ contained in the image of the canonical section $\sigma : S
\to Y$, $E$ is the elliptic fiber and $N_5$ is a non-negative integer.
Note that $\Xi$ can be naturally identified to an effective curve on $S$.
Then we impose

$III)$ The connected components of $\Xi$ are smooth irreducible $(-1)$ curves on $S$.

\noindent
We will discuss the relation between this condition and \kah\ moduli stabilization
in subsection \S 3.10.

Ideally one would like to construct both $V_1,V_2$ in terms of spectral data \refs{\FMW,\FMWII,\Di},
but this approach may be too restrictive, given the constraints $(I)-(III)$ above.
In fact $V_2$ will be indeed constructed in terms of spectral data, but not $V_1$,
which will be constructed by extensions.
Let us first review some aspects of the spectral cover
construction.

\subsec{Spectral covers}

There is a one-to-one correspondence between bundles $V\to Y$, flat and semistable along
the elliptic fibers, and spectral data $(\CC,\CN)$. $\CC$ is an effective divisor on $Y$ flat
over $S$, and $\CN$ is a torsion free rank one sheaf on $\CC$ \refs{\FMW,\FMWII,\Di}.
In order to construct the bundle $V$ in terms of spectral data $(\CC,\CN)$, first take the
fiber-product $$T=\CC\times_S Y\subset Y\times_S Y.$$
Let $p_\CC, p_Y$ denote the canonical projections onto the two factors
and $\pi_T : T \to S$ denote the projection to $S$.
Note that we have three natural divisor classes $\Delta, \sigma_1, \sigma_2$
obtained by restriction from
$Y\times_S Y$. $\Delta$ is the restriction of the diagonal, and
$\sigma_1, \sigma_2$ are restrictions of the canonical sections
$\sigma_1=\sigma \times_S Y$, $\sigma_2=Y\times_S \sigma$. The relative Poincar\'e line bundle
on $Y\times_S Y$ is defined by
\eqn\poincare{
\CP = \CO(\Delta - \sigma_1 -\sigma_2) \otimes \pi^*_SK_S
}
where $\pi_S:Y\times _S Y \to S$ denotes the natural projection.
Then $V$ is given by the push-forward
\eqn\bundleA{
V = p_{Y*}\left(p_\CC^*\CN \otimes \CP\big |_{T} \right)
}
The topological invariants of $V$ are determined by the linear equivalence class of $\CC$ and the Chern
class
of $\CN$. In particular, if the class of $\CC$ is of the form
\eqn\spectralA{
\CC = n \sigma + \pi^* \eta}
with $\eta$ a divisor class on $S$, $V$ will have rank $n$.

Suppose $\CC$ is irreducible and meets the section $\sigma$ along an effective curve $F\subset S$.
Then we have $\CO_T(\sigma_1) \simeq p_\CC^*\CO_\CC(F)$. Moreover, the only generic line bundles
on $\CC$ are $\CO_\CC(F)$ and line bundles pulled back from $S$. Therefore a generic bundle $V$ will
be of the form
\eqn\bundleB{
V=p_{Y*}\left(\CO_T(\Delta -\sigma_2) \otimes p_\CC^*(\CO_\CC(-aF)) \otimes \pi_T^*\CM\right)}
for an integer $a$, and a line bundle $\CM$ on $S$.
Following \refs{\FMWII}, we will denote by $V_{n,a}[\CM]$ a rank $n$ bundle of the form \bundleB\
and by $V_{n,a}$ a bundle of the form \bundleB\
with $\CM\simeq \CO_S$. Note that $V_{n,a}[\CM] \simeq V_{n,a} \otimes \pi^*\CM$.

The Chern character of a bundle of the form $V_{n,a}[\CM]$ is given by
\refs{\FMWII} (Thm. 5.10)
\eqn\bundleC{
\hbox{ch}\left(V_{n,a}[\CM]\right) =
\left[e^{-\eta} \left({1-e^{(a+n)c}\over 1-e^{c}}\right) -{1-e^{ac}\over 1-e^{c}}+
e^{-\sigma}(1-e^{-\eta})
\right] \cdot e^{c_1(\CM)}}
where $c=\pi^*c_1(S)$. In particular we have
\eqn\bundleD{
\eqalign{
& \hbox{ch}_1\left(V_{n,a}[\CM]\right)=
-(n+a-1)\eta +\left[an+{n^2-n\over 2}\right]c_1(S)+nc_1(\CM)\cr
& \hbox{ch}_3\left(V_{n,a}[\CM]\right)={1\over 2}(\sigma^2\eta+\sigma\eta^2)-\sigma \eta c_1(\CM).\cr}}
For future reference, note that for a bundle of the form $V_{n,a}[\CM]$, the spectral line bundle
$\CN$ in \bundleA\ is of the form
\eqn\spectralD{
\eqalign{
\CN & \simeq \CO_\CC((1-a)F) \otimes \pi^*(K_S^{-1} \otimes \CM)\cr
& \simeq \left(\CO_Y((1-a)\sigma) \otimes \pi^*(K_S^{-1} \otimes \CM)\right)\big |_{\CC}.\cr}}

According to \refs{\FMWII} (Thm.7.1) if $\CC$ is irreducible it follows that  $V$ is stable with respect
to a polarization of the form
\eqn\pol{J = \epsilon J_0 + \pi^* H}
where $J_0$ is a fixed ample class on $Y$, $H$ is an ample class on $S$, and $\epsilon$ is a
sufficiently small positive number. Sufficient criteria for the spectral cover to be irreducible have been
formulated in \refs{\DOPWI,\OPP,\DLOPI,\DLOPII,\DOPWII}. They show that
$\CC$ is irreducible if

$i)$ $|\eta|$ is a base point free linear system on $S$, and

$ii)$ $\eta-nc_1(S)$ can be represented by an effective curve on $S$.

\noindent
By Bertini's theorem, the first criterion is satisfied if $\eta$ is ample on $S$, which
in turn amounts to the numerical condition

$i')$ $\eta \cdot \zeta \geq  0$ for all generators $\zeta$ of the Mori cone
of $S$.

\noindent
We will make heavy use of these criteria later in this section. In order to construct the hidden sector
bundle we have to invoke a generalization of the above formalism dealing with reducible spectral covers.

\subsec{Reducible spectral covers and extensions}

Let assume now that $\CC$ is a reducible spectral cover with two smooth reduced irreducible components
\eqn\spectralC{
\CC = \CC' + \CC''}
intersecting along a smooth irreducible curve $C=\CC'\cap \CC''$.
The two components are equipped with spectral line bundles $\CN',\CN''$ so that the restrictions
$\CN'\big |_C, \CN''\big |_C$ are isomorphic. Note that if we choose an isomorphism $\phi :
\CN'\big |_C\to  \CN''\big |_C$, the data $(\CN',\CN'',\phi)$ determines a line bundle $\CN$ on the
reducible spectral cover $\CC$.

Following \refs{\BJPS} (\S 5.1) to any such spectral data $(\CC,\CN)$
we can associate a set $(D,Q)$ of gluing data.
$D$ is a vertical divisor on $Y$ constructed by projecting $C$ to the base $S$,
and then taking the inverse image, $D=\pi^{-1}(\pi(C))$. Let $\CD$ denote the intersection of the fiber
products $\CC'\times_S Y$ and $\CC''\times_S Y$ in $Y\times_S Y$. Note that we have natural projections
$p_\CD : \CD \to D$ and
$\pi_\CD : \CD \to C$. We can use $\pi_\CD$ to pull back the restriction of $\CN'$ (or equivalently
$\CN''$) to $\CD$, obtaining a line bundle on $\CD$ which will be denoted by $\pi_\CD^*\CN$.
Then $Q$ is defined by the following push-forward formula
\eqn\gluingA{
Q = p_{\CD*}\left(\pi_\CD^*\CN \otimes \CP\big |_\CD\right).}

According to \refs{\BJPS} (\S 5.1) the bundle $V\to Y$ corresponding to
the spectral data $(\CC,\CN)$ is given by the following elementary
modification
\eqn\elmodA{
0\to V \to V'\oplus V'' \to Q\to 0}
where $Q$ is regarded as a torsion sheaf on $Y$ supported on $D$.

We will be interested in a special case of this construction when $\CC'=\sigma$ with multiplicity $1$,
and $\CC''$ is a smooth irreducible component which intersects $\sigma$ along a smooth curve
$C$. Note that in this case, $C$ is identical to the curve $F$ for $\CC''$, which was
introduced below \spectralA.\ Moreover, we take
both $\CN'$ and $\CN''$ to be restrictions of a line bundle $\CN$ on $Y$ of the form
\spectralD.\ In particular,
$\CN' \simeq K_S^{(-a)}\otimes \CM$.
This case was treated in detail in \refs{\FMWII} (\S 5.7). The bundle $V'$ is isomorphic to
$\pi^*(K_S^{-a} \otimes \CM)$ and the elementary modification \elmodA\ reduces to
\eqn\elmodB{
0 \to V \to \pi^*(K_S^{-a} \otimes \CM)\oplus V'' \to \pi^*(K_S^{-a} \otimes \CM)|_{D} \to 0.}

As explained in \refs{\BJPS} (\S 5.1), the elementary modification \elmodB\ has moduli
parameterized by the linear space
\eqn\elmodC{
\hbox{Ext}^1(V'(-D),V'')\oplus \hbox{Ext}^1(V'',V'(-D))}
up to $\IC^*$ identifications. The first
direct summand in \elmodC\ parameterizes extensions
of the form
\eqn\extensionA{
0 \to V'(-D) \to V \to V''\to 0}
while the second direct summand in \elmodC\ parameterizes extensions of the form
\eqn\extensionB{
0 \to V'' \to V \to V'(-D)\to 0.}
In particular this shows that the bundle $V$ is a deformation of the direct sum
$V'(-D)\oplus V''$. For future reference note that
\eqn\pullbackform{
V'(-D) \simeq \pi^*(K_S^{-a} \otimes \CM\otimes \CO_S(-F)).}

Summarizing this discussion, it follows that we can construct a more general class of
bundles on $Y$ associated to reducible spectral covers by taking extensions.
In order for this construction to be useful in physical applications, we would like to
have a stability criterion for bundles of this form. Fortunately, such a criterion has been
formulated in \refs{\DOPWI,\DLOPI,\DLOPII,\DOPWII}. Given an extension of the form
\eqn\extensionC{
0\to E' \to E\to E''\to 0}
where $E',E''$ are stable bundles corresponding to irreducible spectral covers,
$E$ is stable if

$a)$ the extension \extensionC\ is not split, and

$b)$ $\mu_J(E') < \mu_J(E)$, where the slope $\mu_J$ of a bundle $E$ is defined by
$$\mu_J(E)= {c_1(E)\cdot J^2\over {\hbox{rk}(E)}}.$$

\noindent
Here by stability we mean stability with respect to a polarization of the form \pol\
with sufficiently small $\epsilon>0$.

This is all the formal machinery we will need, so we can turn to the explicit construction
of bundles.

\subsec{The GUT bundle}

Let us first construct the GUT $SU(5)$ bundle $V_2$. The spectral cover construction suffices
in this case. From now on we take the base $S$ to be the del Pezzo surface $dP_8$.
Pick a spectral cover $\CC$ in the linear system $$|5\sigma -6\pi^* K_S|$$
which implies $\eta=6c_1(S)$.
Take $V_2$ to be a bundle of the form $V_{5,1}[K_S^{-3}]$. Then, using formulas
\bundleD\ a straightforward computation shows that
\eqn\bundleE{
\hbox{ch}_1(V_2)=0,\qquad \hbox{ch}_3(V_2)=-3}
in agreement with \generations.\
Note that the stability criteria $(i')$, $(ii)$ formulated below \pol\ are automatically
satisfied because $\eta=6c_1(S)$ is very ample on $S$.
Therefore $V_2$ is stable and has three generations.

\subsec{The hidden sector bundle}

In this case we have not been able to find an irreducible spectral cover construction satisfying
conditions $(I)-(III)$. We will however construct an $SU(5)$ bundle $V_1$ with the
required properties
using a reducible spectral cover and extensions as explained in \S 3.5.

Let us consider a reducible spectral cover of the form $\spectralC$ with $\CC'=\sigma$
and $\CC''$ a smooth irreducible divisor in the linear system $$|4\sigma -\pi^*(6K_S+\Gamma)|$$
where $\Gamma$ is a smooth irreducible $(-1)$ curve on $S$. For example we can take
$\Gamma$ to be any generator of the Mori cone of $S$.
Recall \refs{\DHOR} that the Mori cone of $S=dP_8$ is generated by the $240$
$(-1)$ curve classes
\eqn\Morigen{
\eqalign{
& e_i,\ h-e_i-e_j,\ 2h-e_i-e_j-e_k-e_l-e_m,\cr
& 3h-2e_i-e_j-e_k-e_l-e_m-e_n-E_o,\cr
& 4h-2(e_i+e_j+e_k) -\sum_{s=1}^5 e_{m_s},\cr
& 5h-2\sum_{s=1}^6e_{m_s}-e_k-e_l,\
6h-3e_i-2\sum_{s=1}^7 e_{m_s},\cr}}
where $h$ is the hyperplane class and $e_i$, $i=1,\ldots, 8$ are the exceptional curve
classes. The indices $i,j,k,l,m,n,o,m_s$ in \Morigen\
are pairwise distinct and take values from $1$ to $8$.
We will denote by $\eta''=-6K_S-\Gamma=6c_1(S)-\Gamma$.

One can check that that $\CC''$ is  indeed
irreducible using the criteria $(i')$, $(ii)$ below equation \pol.\
We have to check that $\eta''\cdot \zeta\geq 0$ for any of the generators $\zeta$ listed in
\Morigen.\ This follows by direct computations. Moreover, if $\Gamma$ is any of the Mori cone
generators, one can check that the class $\eta''-4c_1(S)= 2c_1(S)-\Gamma$ is again a generator.
Therefore $\eta''-4c_1(S)=F$ is an effective $(-1)$ curve class on $S$.

We pick the spectral line bundles $\CN',\CN''$ to be the restrictions of a line bundle
of the form
\eqn\lineA{
\CN = \CO_Y((1-a)\sigma) \otimes \pi^*(K_S^{-1}\otimes \CM)}
to $\CC'$ and respectively $\CC''$.
Then we find
\eqn\lineB{
\CN' = K_S^{-a} \otimes \CM, \qquad
\CN''= \CO_{\CC''}((1-a)F) \otimes \pi^*(K_S^{-1}\otimes \CM).}
The bundles $V',V''$ determined by the spectral data $(\CC',\CN')$, $(\CC'',\CN'')$ are
\eqn\twobundles{
V'\simeq \pi^*(K_S^{-a} \otimes \CM), \qquad
V'' \simeq V_{4,a}[\CM]}
where the bundles $V_{n,a}[\CM]$ have been defined below equation \bundleB.\
Note that both $V', V''$ are stable with respect to a polarization of the form \pol\
since $\CC',\CC''$ are irreducible.

The rank $5$ bundle $V$ associated to the reducible spectral data $(\CC,\CN)$
is determined by an elementary modification of the form \elmodB.\ Let us first compute the
Chern classes of $\ch_1(V),\ch_3(V)$ using formulas \bundleD.\ For the purpose of this
computation, we may assume that $V_5$ is a direct sum $V'(-D) \oplus V''$ since the Chern classes
do not change under deformations. Then we obtain
\eqn\chernclsA{
\eqalign{
\ch_1(V) & = \pi^*\left[(5a+6)c_1(S)-(a+3)\eta''+5c_1(\CM)-F\right]\cr
\ch_3(V) & = \half \eta''(\eta''-c_1(S)-2c_1(\CM)),\cr}}
where we have used the isomorphism $\CO_Y(-D) \simeq \pi^*\CO_S(-F)$.
Substituting $\eta'' = 6c_1(S)-\Gamma$ and $F=2c_1(S)-\Gamma$ in \chernclsA,\ we obtain
\eqn\chernclsB{
\eqalign{
\ch_1(V) & = 5c_1(\CM) -(a+14)c_1(S) +(a+4)\Gamma\cr
\ch_3(V) & = \half( 18-12c_1(S)\cdot c_1(\CM) + 2\Gamma \cdot c_1(\CM)).\cr}}
Note that choosing
\eqn\choice{
a=-4,\qquad  \CM \simeq K_S^{-2},}
we obtain
\eqn\chernclsC{
\ch_1(V)= 0,\qquad \ch_3(V)=-1}
in agreement with the condition $(I)$.

Next we have to check stability using criteria $(a)$, $(b)$ in the previous subsection. Let us
first compute the extension moduli \elmodC.\ We start with
$$
\hbox{Ext}^1(V'',V'(-D))=H^1((V'')^\vee \otimes V'(-D)).
$$
This cohomology group can be computed using the Leray spectral sequence
\eqn\lerayA{
H^p(S,R^q\pi_*((V'')^\vee \otimes V'(-D)))\Rightarrow H^{p+q}((V'')^\vee \otimes V'(-D)).}
Since $V'(-D)$ is pulled back from $S$ according to equation \twobundles,\
the left hand side of \lerayA\ can be simplified to
\eqn\lerayB{
H^p(S, R^q\pi_*(V'')^\vee\otimes K_S^{-a} \otimes \CM \otimes \CO_S(-F)).}
Note that only terms with $(p,q)=(0,1),(1,0)$ can occur in the computation
of $H^1$. The direct images $R^q\pi_* (V'')^\vee$ for a stable spectral cover bundle
have been computed for example in \refs{\DI} (\S 2.1) We have
\eqn\dirimages{
R^0\pi_*(V'')^\vee =0, \qquad
R^1\pi_*(V'')^\vee \simeq K_F \otimes \CN^{-1}|_{F} \otimes K_S^{-1}|_F.}
Note that $R^1\pi_*(V'')^\vee$ is a torsion sheaf on $S$ supported on $F$. Therefore we are left
with
\eqn\lerayC{
\eqalign{
H^1((V'')^\vee \otimes V'(-D))& \simeq H^0(F, K_F \otimes \CN^{-1}|_{F} \otimes K_S^{-1-a}|_F \otimes
\CM|_F \otimes \CO_F(-F)).\cr}}
Equation \lineA\ implies
$$
\CN^{-1}|_F \simeq K_S^{a}|_F \otimes \CM^{-1}|_F.
$$
Substituting this relation in \lerayC\ we obtain
\eqn\lerayD{
H^1((V'')^\vee \otimes V'(-D)) \simeq H^0(F, K_F \otimes K_S^{-1}|_F \otimes \CO_F(-F)).}
Since $F\simeq \IP^1$ is a smooth $(-1)$ rational curve on $S$ we have
$$
K_S^{-1}|_F \simeq \CO_{\IP^1}(1),\qquad \CO_F(-F) \simeq \CO_{\IP^1}(1),\qquad K_F \simeq \CO_{\IP^1}(-2).
$$
Therefore \lerayD\ reduces to
\eqn\lerayE{
H^1((V'')^\vee \otimes V'(-D)) \simeq H^0(\IP^1, \CO_{\IP^1}) =\IC.}
This shows that up to isomorphism we have a unique nontrivial extension of the form
\eqn\extensionD{
0\to V'(-D) \to V \to V''\to 0.}
One can similarly compute
$$
\hbox{Ext}^1(V'(-D),V'') \simeq H^0(\IP^1,\CO_{\IP^1}(-2))=0,
$$
therefore there are no nontrivial extensions of the form
$$
0\to V'' \to V \to V'(-D)\to 0.
$$
In conclusion, taking into account \choice,\ we are left with a unique bundle $V_2$
defined by the unique  nontrivial extension
\eqn\extensionE{
0 \to  \pi^*(K_S^{2}\otimes \CO_S(-F)) \to V_2 \to V_{4,-4}[K_S^2]\to 0.}

According to criterion $(b)$ stated below \extensionC,\ in order to show that $V$ is stable, it suffices
to check that
$$ \mu_J(\pi^*(K_S^{2}\otimes \CO_S(-F))) < \mu_J(V).$$
Since $\ch_1(V_2) =0$, we have $\mu_J(V_2)=0$. For a polarization of the form \pol\ we have
$$
\mu_J(\pi^*(K_S^{2}\otimes \CO_S(-F))) = \pi^*(-4c_1(S)+\Gamma) \cdot (\epsilon^2 J_0 + 2\epsilon\pi^* H).
$$
Note that $(4c_1(S)-\Gamma)$ is an effective curve class on $S$, and $H$ is ample on $S$, therefore
$$
(-4c_1(S)+\Gamma)\cdot H < 0.
$$
Then we can satisfy the criterion by taking $\epsilon >0$ sufficiently small.
\vfill\eject

\subsec{Heterotic fivebranes}

In order to complete the description of the model, we have to compute the heterotic fivebrane class
$\Lambda$ and check that it is effective. Using formula \bundleC,\ it is straightforward to compute
$$
\eqalign{
\ch_2(V_1)& = \pi^*\ch_2(K_S^{2}\otimes \CO_S(-F))+\ch_2(V_{4,-4}[K_S^2])\cr
          & = 5E - \sigma \pi^*(6c_1(S)-\Gamma),\cr
\ch_2(V_2)& = 5E -6\sigma \pi^*c_1(S),\cr}
$$
where $E$ is the class of the elliptic fiber on $Y$.
Therefore we obtain
$$
c_2(V_1)+c_2(V_2)=\sigma \pi^*(12c_1(S)-\Gamma) - 10 E.
$$
The second Chern class of $Y$ is
$$
\eqalign{
c_2(Y) & = 12\sigma \pi^*c_1(S) + (c_2(S)+11c_1(S)^2)E\cr
& = 12\sigma \pi^*c_1(S) + 22E.\cr}
$$
Therefore equation \anomaly\ yields
\eqn\fivebrane{
\Lambda = \sigma \pi^*\Gamma + 32 E,}
which is effective.
Note that in addition to vertical fivebranes wrapping the elliptic fiber, we have a
horizontal fivebrane wrapping the $(-1)$ curve $\Gamma$ in the base.

\subsec{F-theory interpretation}

We conclude this section with a brief discussion of the F-theory dual models. Let us first recall
some aspects of heterotic F-theory duality following \refs{\MVII,\Fmore,\Aiii} and \refs{\FMW} (\S 6.1).
Suppose we have a heterotic model given by two stable bundles $V_{1},V_2$ on a smooth Weierstrass model
$\pi:Y \to S$ corresponding to smooth irreducible spectral covers $\CC_1,\CC_2$. First we assume that the
anomaly cancellation condition \anomaly\ is satisfied in the absence of horizontal fivebranes, that is the
class $\Xi$ introduced below \anomaly\ vanishes. This means that the two classes $\eta_1,\eta_2$ associated
to the spectral covers satisfy
\eqn\anometa{
\eta_1+\eta_2 = 12 c_1(S).}
Let us write
\eqn\etaclasses{
\eta_1= 6c_1(S) -t, \qquad
\eta_2=6c_1(S) + t,}
where $t$ is a divisor class on $S$. Let $\CT$ be a line bundle on $S$ with $c_1(\CT)=-t$.

For our purposes it suffices to consider take the structure groups of $V_1,V_2$ to be $SU(n_1)$ and respectively
$SU(n_2)$. The spectral cover of an $SU(n)$ bundle is determined \refs{\FMW} by $n$ sections $a_k$ of the
line bundles
\eqn\spectralsections{K_S^{k}\otimes \CO_S(\eta)\simeq K_S^{-6+k}\otimes \CT,\qquad
k=0,2,\ldots, n.}

The dual F-theory model is an elliptic fibration $X\to P$ with a section over a base $P$, where $P$ is a $\IP^1$
bundle over $S$. More precisely, $P$ is the projectivization of the rank two bundle of the form $\CO_S \oplus \CT$.
Note that $P$ has two canonical sections $S_0,S_\infty$ with normal bundles
\eqn\normbundles{
N_{S_0/P} \simeq \CT, \qquad
N_{S_\infty/P} \simeq \CT^{-1}.}
The canonical class of $P$ is
$$
K_P = -S_0-S_\infty+p^*(K_S)
$$
where $p:P\to S$ denotes the canonical projection.
The fourfold $X$ is a Weierstrass model of the form
\eqn\weierstrass{
y^2=x^3-fx-g}
where $f,g$ are sections of $K_P^{-4}$ and respectively $K_P^{-6}$. The discriminant of the elliptic fibration is
given by
\eqn\elldiscr{
\delta = 4f^3-27g^2.}
We will denote by capital letters $F,G,\Delta$ the zero divisors of $f,g,\delta$ on $P$.

The heterotic bundles $V_1,V_2$ correspond to $ADE$ degenerations of the elliptic fibration $X\to P$ along the sections
$S_0$, $S_\infty$. The discriminant $\Delta$ decomposes into three components
\eqn\discrcomp{
\Delta = \Delta_0 + \Delta_\infty + \Delta_n}
where $\Delta_0$, $\Delta_\infty$ are multiples of $S_0$, $S_\infty$ and $\Delta_n$ denotes the nodal component.

The heterotic bundle moduli are encoded in the complex structure moduli of $X$.
For concreteness, let us consider the first
bundle $V_1$ which corresponds to an $ADE$ degeneration along $S_0$; $V_2$ follows by analogy replacing $S_0$
with $S_\infty$
and $\CT$ with $\CT^{-1}$ in the following. The complex structure of the elliptic fibration in a neighborhood of
$S_0$ is
captured by a hypersurface equation of the form
\eqn\elldegA{
y^2=x^3-fx-g}
in the total space of the rank three bundle $\CT\oplus (\CT^2\otimes K_S^{-2}) \oplus (\CT^3\otimes K_S^{-3})$.
Let $s$ denote
a linear coordinate on the total space of the line bundle $\CT\to S$. Then $f,g$ in \elldegA\ have
expansions of the form
\eqn\expansions{
\eqalign{
f & = f_0+f_1s+f_2s^2 + \ldots \cr
g &  = g_0 +g_1s +g_2s^2 + \ldots \cr}}
where $f_i$ are sections of $K_S^{-4}\otimes \CT^{4-i}$, $i=0,\ldots,4$ and $g_j$ are sections of
$K_S^{-6}\otimes \CT^{6-j}$,
$j=0,\ldots,6$. The duality map relates the nonzero sections $f_i,g_j$ to the sections $a_k$ of the
line bundles \spectralsections\
which determine the heterotic spectral cover.

This picture is valid as long as the spectral cover is irreducible and there are no horizontal heterotic
fivebranes. Horizontal
heterotic fivebranes correspond to blow-ups in the base of the F-theory elliptic fibration
\refs{\DR,\ACiii}. More
specifically, suppose we have a single heterotic fivebrane wrapping a smooth curve $\Xi$ contained in the
section of the Weierstrass model
$Y$. In F-theory, this is represented by performing a blow-up of the base $P$ along a curve isomorphic
to $\Xi$ contained in a section $\CS$ of the $\IP^1$ fibration $p:P\to S$.
We will call this curve $\Xi$ as well since the distinction will be clear from the context.
This is the four-dimensional counterpart of the more familiar six-dimensional F-theory picture for small instantons
developed in \refs{\MVII,\ATasi,\Apoint,\AM}.
The vector bundle degenerations associated to heterotic fivebranes on Calabi-Yau threefolds
have been studied in \refs{\Curio,\BDO,\OPP}.

Let $\wP$ denote the total space of the blow-up, and let ${\widetilde \CS}$ denote the proper transform
of $\CS$ in $\wP$. Note that both
$\CS$ and ${\widetilde \CS}$ are naturally isomorphic to $S$. Then the normal bundle of ${\widetilde \CS}$
in $\wP$ is determined by
\eqn\normbundle{
N_{{\widetilde \CS}/\wP}\simeq N_{\CS/P} \otimes \CO_S(-\Xi).}

Let us now construct the F-theory dual of our model. The line bundle $\CT$ is isomorphic to $\CO_S$ in our case,
and we have a horizontal
heterotic fivebrane wrapping the $(-1)$ curve $\Gamma$. Therefore the F-theory base is a blow-up of
$$
P=\IP^1 \times S
$$
along a curve isomorphic to $\Gamma$ contained in a section $\CS$. According to equation
\normbundle,\ $N_{{\widetilde \CS}/\wP}\simeq
\CO_S(-\Gamma)$, hence ${\widetilde \CS}$ is rigid in $\wP$. We will take
${\CS} = S_0$ in the following.

We have two $SU(5)$ bundles $V_1, V_2$ on $Y$. Recall that the bundle $V_2$ admits an irreducible
spectral cover description with
$\eta_2=6c_1(S)$. According to the duality map reviewed above, $V_2$ corresponds to an $A_4$
degeneration of the fourfold $X$ along
a section $\Sigma$ of the fibration $\wP\to S$ with trivial normal bundle.
Such a section $\Sigma$ is in fact the proper transform of any
section of $P$ over $S$ distinct from $S_0$. Note that $\Sigma$ moves in a one dimensional linear system,
and it can degenerate to a reducible
divisor of the form $\wS_0+E$, where $E$ is the exceptional divisor of the blow-up map $\wP\to P$.

There is one more condition on the $A_4$ degeneration along $\Sigma$, namely it has to be split \refs{\Fmore}.
This means that the degeneration
has no monodromy along curves in $\Sigma$. The conditions for an $A_4$ singularity to be split have been derived
in \refs{\Fmore}. The
expansions \expansions\ have are truncated to
\eqn\expansionsB{
\eqalign{
f &= s^2 f_2 + \ldots + s^4f_4 \cr
g & = s^3 g_3 + \ldots + s^6g_6\cr}}
where $f_i$ are sections of $K_S^{-4}$, $i=2,\ldots,4$, and $g_j$ are sections of $K_s^{-6}$, $j=3,\ldots,6$ since
$\CT\simeq \CO_S$. The split
condition requires $f_2,f_4$ and $g_3,g_4,g_6$ to be written as polynomial functions of a a smaller set of sections
$$h\in H^0(K_S^{-1}),
\quad H\in H^0(K_S^{-2}),\quad  q\in H^0(K_S^{-3}).$$ Therefore the complex structure moduli of a split $A_4$
degeneration along the section
$\Sigma$ are controlled by the coefficients
$$
h\in H^0(K_S^{-1}),\quad  H\in H^0(K_S^{-2}),\quad  q\in H^0(K_S^{-3}), \quad
f_3\in H^0(K_S^{-4}),\quad  g_6\in H^0(K_S^{-6}).
$$
These sections are in one-to-one correspondence with the sections \spectralsections\ which determine the spectral
cover $\CC_2$.

The duality map for bundles with reducible spectral covers, or more generally bundles constructed by extensions,
is more subtle and
not completely understood. Here we
are only interested in reducible spectral covers of the form $\CC= \sigma + \CC''$,
and extensions of the
form \extensionE.\ According to \refs{\BJPS} (\S 5.1) in this case, the moduli of the rank four bundle $V''$
corresponding to $\CC''$
are mapped to complex structure deformations of the fourfold. The extension moduli are related to expectation
values of D7-D7 fields which
span nongeometric branches in F-theory. This correspondence is not understood in detail at the present stage,
but the current understanding
suffices for our purposes.

Recall that $\CC''$ is an irreducible spectral cover with $\eta'' = 6c_1(S) -\Gamma$, therefore
$\CT \simeq \CO_S(-\Gamma)$. This is the
normal bundle of the proper transform $\wS_0$ in $\wP$, hence we will have a fourfold degeneration
along the rigid section $S_0$. The spectral
cover moduli \spectralsections\ are  parameterized  this case by
\eqn\spectralmod{
\eqalign{
& a_0\in H^0(K_S^{-6}(-\Gamma)), \qquad a_2\in H^0(K_S^{-4}(-\Gamma))\cr
& a_3\in H^0(K_S^{-3}(-\Gamma)), \qquad a_4 \in  H^0(K_S^{-2}(-\Gamma)).\cr}}
Using the results of \refs{\Fmore} (\S 4.5) one can check that these are precisely the complex structure moduli
of a split $D_5$ singularity
along the section $\wS_0$ \refs{\Fmore}. So naively we seem to obtain a $SO(10)$ gauge group in F-theory. However,
this is not true, since
we have not taken into account the D7-D7 strings localized on the intersection between $\wS_0$ and the nodal
component of the discriminant.
The expectation values of these fields should be related to extension moduli in the heterotic model \refs{\BJPS}.
In our case, we have no extension
moduli, as shown at the end of \S 3.6. We can construct only a direct sum
bundle which is unstable or a non-split
extension, which is stable. Note
that this behavior is not solely determined by the spectral cover $\CC''$. The line bundle $\CN$ also plays a crucial
role in the extension moduli computation.
The choice of line bundle is expected to be related to the background flux $G$
in F-theory \refs{\Curio,\CurD,\AC}.

Given all this data, we propose the following interpretation of the F-theory
dual. Naively, the gauge symmetry seems to be $SO(10)$, but with
the present combination of geometry and flux, the D7-D7 strings localized on
the intersection are tachyonic.
Therefore they must condense
spontaneously breaking the $SO(10)$ gauge symmetry on the D7-brane wrapping
$\wS_0$ to $SU(5)$.
The tachyonic nature of the D7-D7 strings is related to the fact that the
direct sum bundle is unstable,
therefore breaks supersymmetry at tree level in the heterotic model. It would be
very interesting to understand this correspondence in more detail, but we leave this for future work.

Summarizing this discussion, we conclude that one can construct an
F-theory dual to the previous heterotic model.
The geometry of the F-theory
fourfold is an elliptic fibration over a blow-up $\wP$ of
$\IP^1\times S$ and the GUT $SU(5)$ bundle corresponds to a
split $A_4$ degeneration
along a movable section of $\wP$ over $S$. The hidden $SU(5)$ bundle corresponds to
a split $D_5$ degeneration along a
rigid section, and the
interplay of geometry and flux leads to tachyon condensation in F-theory, breaking $SO(10)$ to $SU(5)$.

An important point for us is that the resulting F-theory fourfold has complex structure
deformations which bring the movable section supporting the GUT $SU(5)$ arbitrarily close
to the fixed section supporting the hidden $SU(5)$. This allows us to tune the messenger masses to be small
enough to arrange for gauge mediation as the dominant source of SUSY breaking in the observable sector:
different values of the fluxes which stabilize the complex structure moduli lead to a wide range of possibilities for $M$.

Since this point is slightly subtle, let us provide more details.
First note that this question reduces to the analogous problem in
eight dimensions\foot{We thank T. Pantev for clarifying discussions on these points.}
since the F-theory elliptic fibration is a (blow-up of a) direct product
$P=\IP^1 \times S$. This means it suffices to consider an eight dimensional F-theory compactification on an
elliptically fibered K3 surface with a section. We will be interested in a subspace of the
moduli space where the K3 surface has one Kodaira fiber of type $I_5$ (corresponding to an $A_4$
singularity) and another Kodaira fiber of type $I_1^*$ (corresponding to a $D_5$ singularity) in addition to $12$ Kodaira
fibers of type $I_1$. In this case one can construct a one parameter family of K3 surfaces so that
the generic surface has singular fibers $I_5+I_1^*+12 I_1$ as above and the central fiber
has singular fibers of type $I_3^*+15 I_1$. (The singular fiber $I_3^*$ corresponds to a $D_7$ singularity.)
Locally, this is a collision of $I_5$ and $I_1^*$ singular fibers; it is constructed explicitly in \S 11 of \Miranda.\

\ifig\qpoly{Reflexive polyhedron describing the family of K3 surfaces. The color code is as follows: 
magenta is associated with the
elliptic fiber of the $Spin(32)/\IZ_2$ fibration, while red corresponds to the associated extended Dynkin diagram;
green is associated with the elliptic fiber of the $E_8\times E_8$ fibration, while blue corresponds to the respective 
extended Dynkin diagrams.}{\epsfxsize2.0in\epsfbox{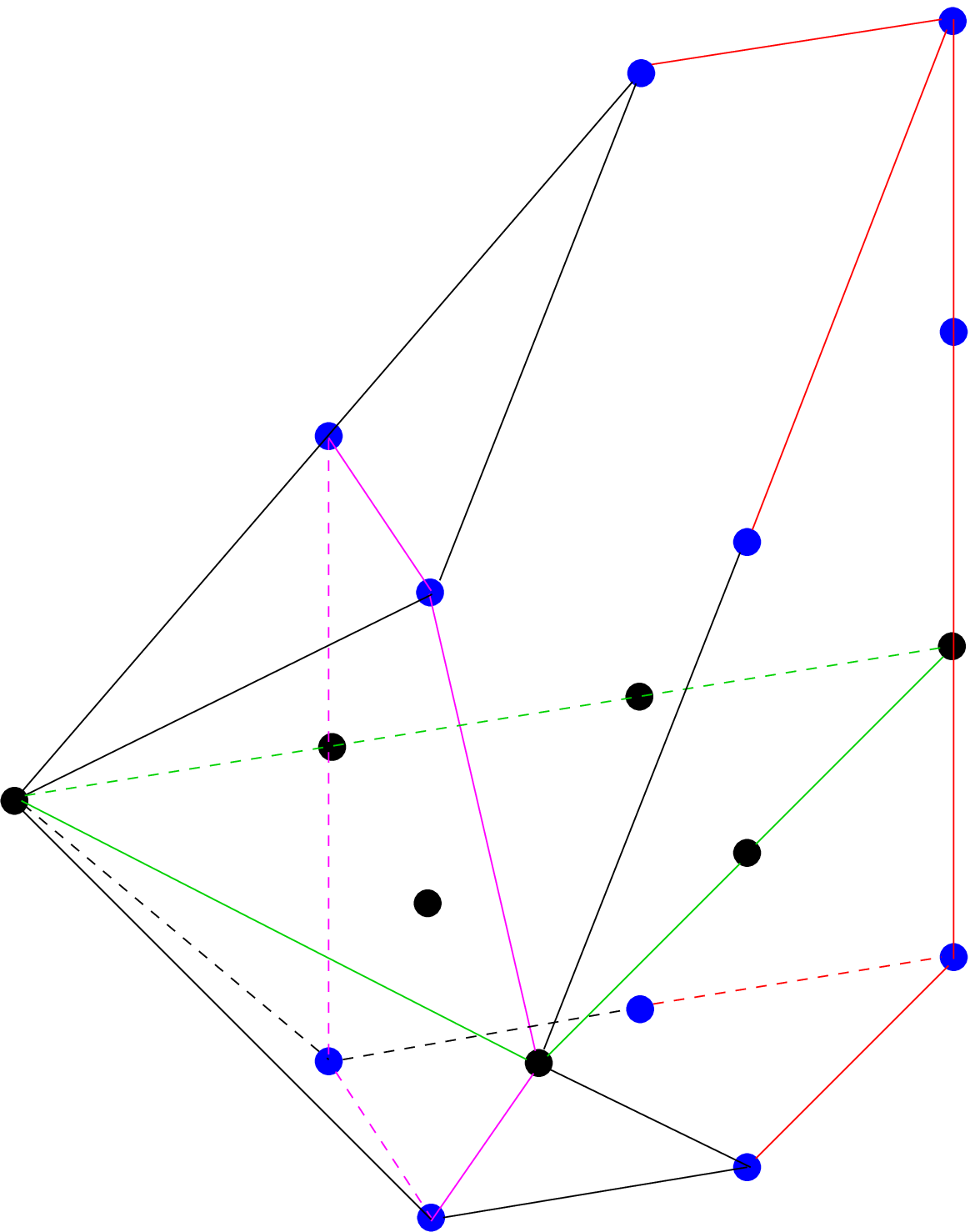}}

In fact we can give an alternative argument based on a presentation of these
K3 surfaces as hypersurfaces in toric varieties; this is based on a beautiful observation due to Candelas and Skarke \refs{\CSi}.
We represent the corresponding toric polyhedron below. Note that the K3 surfaces develop a $D_7$ singularity along a certain subspace
of the moduli space. To see this, note that the polyhedron contains a reflexive subpolyhedron drawn with magenta in fig. 1; the
corresponding torus fibration admits one singular $I_3^*$ fiber. The red edges of the $\nabla$ polyhedron
correspond to the nontrivial pairwise intersections of the exceptional divisors obtained by resolving the $D_7$ singularity. 
Complex structure deformations in the normal directions to this subspace will split the $D_7$
singularity in $A_4+D_5$. To see this, note that the polyhedron contains another reflexive subpolyhedron, which is the elliptic
curve in $\IP^2_{[1,2,3]}$; this is represented with green in fig. 1. The corresponding elliptic fibration has one $D_5$ and one
$A_4$ singularity; the exceptional divisors associated with the resolution of the singularities and the affine components are represented
with blue in fig. 1.

For completeness, note also that the elliptic fibrations with $D_{7}$ singularities admit
two sections. This suggests that our model can be equally well constructed using the $Spin(32)/\IZ_2$
heterotic string, according to \refs{\Apoint}. This is not surprising since it is well known
that the two heterotic models are equivalent when compactified on a two-torus.

In order to put this discussion in proper perspective, note that the picture developed here
does not contradict the more familiar parameterization of complex structure moduli in F-theory.
Usually one fixes the locations of the $A_4$ and respectively $D_5$ singularities at $\infty$ and respectively $0$ using the
$PSL(2,\IC)$ automorphism group of the base $\IP^1$. Then the moduli of the K3
surface are parameterized by deformations of the K3 surface preserving these singularities.
Note however that in this manner one obtains only a parameterization of an open subset
of the moduli space where the two singular points are away from each other. The construction sketched
above yields a parameterization of a different open subset of the moduli space,
centered on the subspace of K3 surfaces with $D_7$ singularities.

Note also that the parameterization commonly used in the literature covers a neighborhood of the
$E_8\times E_8$ semi-stable
degeneration locus in the moduli space \refs{\Apoint,\FMW}. In this region, the F-theory model
admits an alternative description in terms of heterotic M-theory. The region we are interested in
is not near the $E_8\times E_8$ semi-stable degeneration locus, and the model does not admit a
heterotic M-theory interpretation.

\subsec{K\"ahler moduli stabilization}

The typical \kah\ moduli stabilization mechanism in F-theory models relies on
D3-brane instanton effects \Edinst.
The D3-brane instantons which contribute to the
nonperturbative superpotential are classified by arithmetic genus one divisors
$D$ in the resolution of $X$ which project to a surface in the base $P$.
If the base is a $\IP^1$ bundle over a surface $S$, one can show
\refs{\G} that the inverse image of any $(-1)$ curve $C$ in $S$
is such a divisor $D$ which contributes to the superpotential.
In our case, the base is $S=dP_8$, and we can find $240$ $(-1)$ curves
which generate the Mori cone \Morigen.\ Moreover, the base is in fact
a blow-up of $P=\IP^1\times S$ along a curve $\Xi$ lying in a section.
According to \refs{\DR}, the inverse image of the exceptional divisor
$E\subset \wP$ contributes to the superpotential if $\Xi$ is a
$(-1)$ curve as well. This is precisely condition $(III)$ formulated in section \S 3.3,
which is is satisfied in our model.

Finally, the vertical divisors obtained by ruling the exceptional
components of the $D_5$ fiber over the section $S_0$ are rigid and
have arithmetic genus $1$. Therefore, according to \refs{\Edinst},\ they
contribute to the nonperturbative superpotential.
We conclude that this model admits sufficient
contributions to the superpotential such that it is possible to find vacua
with all the \kah\ moduli fixed.

\newsec{Dynamical Supersymmetry Breaking from Quivers}

D3-branes placed at a smooth point in a Calabi-Yau manifold realize a
world-volume gauge theory which, at low energies, flows to $\CN=4$ supersymmetric
Yang-Mills theory.  To get theories with less supersymmetry, one can
place the branes at singular points in the Calabi-Yau: simple examples
include orbifolds \refs{\DM,\orbCFT,\orbtwo} and conifolds \KW.
Suitable classes of singularities may also include collapsed curves;
in such cases, one can sometimes make supersymmetric configurations which
include some number of D5-branes wrapping the collapsed curve (often
called ``fractional branes'').
For instance, the famous Klebanov-Strassler solution \KS\ arises in this way,
by placing D3-branes and wrapped D5-branes at a conifold singularity.

By now, quite a bit has been learned about the classes of so-called
quiver gauge theories which arise from general
configurations of fractional branes at the singularities in Calabi-Yau
moduli space where a divisor collapses to zero size.  A nice review
with references can be found in \Verlinde.
One very interesting insight
which has recently emerged is that quiver theories which preserve SUSY
to all orders in perturbation theory, but break it non-perturbatively,
are easy to find (and may even be generic) \quiverDSB.

This suggests that an easy way to make a model of gauge mediation, may be
to realize the Standard Model on one stack of D-branes, and a quiver theory
which dynamically breaks SUSY on another stack.
By now there are many papers which realize variants of the SM on different
kinds of brane stacks; we will avail ourselves of two
different kinds of constructions.  On the one hand, we can realize
the Standard Model as in
\Verlinde, which realizes the SM using fractional branes in a
(partially) collapsed
$dP_8$.
As long as the DSB quiver is sufficiently close to the SM branes,
the interbrane strings connecting the two stacks will have a mass
$M << M_s$; they will serve as the messengers of gauge mediated supersymmetry
breaking.  On the other hand, we can also arrange for our quiver theory
to arise ``close to'' a GUT D7-brane stack, which is
the F-theory dual to a standard heterotic GUT.

One of the simplest quiver theories which exhibits DSB \quiverDSB\ involves
an $U(3M) \times U(2M) \times U(M)$ gauge group with matter fields
in the
\eqn\mattspec{
({\bf {\overline {3M}}},{\bf 2M},{\bf 1}),
~2\times ({\bf 3M},{\bf 1},{\bf \overline{M}}),~
3\times ({\bf 1},{\bf \overline {2M}},{\bf M})~}
representations (and a suitable tree-level superpotential).  It can be
obtained from fractional branes at a collapsed $dP_1$ as in \quiverDSB.
\foot{We note that the del Pezzo quiver gauge theories may actually have several
branches, with the dynamical supersymmetry breaking vacua being the result of
local analysis on some subset of the branches.  Simpler gauge theories
like the Klebanov-Strassler theory \KS\ already exhibit a very rich
branch structure \DKS.  We thank N. Seiberg for emphasizing this point
to us.}

Therefore, in the next sections, we engineer appropriate collapsed
surface singularities close to: 1) F-theory duals of heterotic GUT models and,
2) avatars of the D3-brane MSSM-like model of \Verlinde. 
In the remainder of this section, we 
discuss some very minor modifications of the example given in \quiverDSB,
which will arise more easily in our geometric engineering.

Before proceeding, we should discuss an important caveat.
As was correctly described in the second paper in \quiverDSB, these theories
have several FI terms associated with the $U(1)$ factors in the 
gauge group.  The $U(1)$'s are anomalous.  The anomalies are cancelled
by a Green-Schwarz mechanism whereby $U(1)$ gauge transformations
are accompanied by shifts of twisted RR axions, and  
the K\"ahler modulus partners of the axions play the role of
field-dependent FI terms \Rabadan. 
Supersymmetry is broken for any fixed finite value of these terms.  
The question of whether supersymmetry is broken once the
FI terms 
become dynamical is a more detailed and subtle one, depending both on
details of the gauge theory which are hard to compute, and details of
the global embedding.  If these dynamical FI terms are not stabilized,
then instead of supersymmetry breaking one finds a run-away to infinity
in field space with no stable vacuum.
We discuss this issue in \S5.3\ after developing more of the
relevant geometry; it is worth
pointing out that there is no analogous issue with the models of \S3.

\subsec{DSB from the $dP_8$ quiver}

We will be mostly interested in compact Calabi-Yau manifolds that admit
$dP_8$ and $dP_5$ singularities (instead of $dP_1$). Here
we briefly explain the elementary point, that this does not hamper us in
using the construction of \quiverDSB. This is in keeping with their
statements about the genericity of the phenomena they discuss.

A three-block exceptional collection for the $dP_8$ singularity is provided
in \S3\ of \Verlinde.  Recall that the middle cohomology of $dP_8$ is spanned
by the hyperplane class $h$ and the exceptional curves $e_i$, $i=1,\ldots,8$ with
\eqn\intnum{h \cdot h = 1, ~e_i \cdot e_j = -\delta_{ij},~h \cdot e_i = 0.}
The canonical class is given by
\eqn\cancl{K_{dP_8} = -3h + \sum_{i=1}^8 e_i.}
A given member $F$ of the exceptional collection can be specified by a charge vector
$({\rm rank}(F),c_1(F),\ch_2(F))$ giving the Chern classes of the sheaf it represents
(the D7, D5 and D3 charge respectively). If one takes $n$ copies of a
given member, one finds a $U(\vert n \vert)$ gauge theory with no adjoint matter.
Given two such sheaves $F_i, F_j$, the spectrum of bifundamentals may be computed in terms
of the Euler character $\chi(F_i,F_j)$.

\noindent The results for $dP_8$ are that there exists an exceptional collection with
$$
\eqalign{
&\ch(F_i) = (1,h - e_i, 0),~i=1,\ldots,4,~~\ch(F_j) = (1,-K_{dP_8} + e_i,0),~j=5,\ldots,8,\cr
&\ch(F_9) = (1,2h - \sum_{i=1}^4 e_i,0),~~\ch(F_{10}) = (3,-K_{dP_8} + \sum_{i=5}^8 e_i,-1/2),\cr
&\ch(F_{11}) = (6,-3K_{dP_8} + 2\sum_{i=5}^8 e_i,1/2).}
$$
The spectrum of bifundamentals is given by
\eqn\bifunds{\eqalign{
&\chi(F_{10},F_i) = 1,~~\chi(F_{11},F_i) = 1 ~{\rm for}~i=1,
\ldots,9,~~\chi(F_{10},F_{11}) = 3.
}}

It was observed in \Verlinde\ that this quiver allows for a simple
construction of a pseudo-realistic MSSM.  To get this toy model,
the multiplicities at the nodes should be chosen to be
\eqn\mult{
n_i=1,~i=1,\ldots,9,~~n_{10}=3,~~n_{11}=-3 ~.
}
In this case,
the charges of the fractional brane add up to the charge of a single D3-brane.

\ifig\quiver{The $dP_8$ quiver associated with the mutated exceptional collection.}
{\epsfxsize3.3in\epsfbox{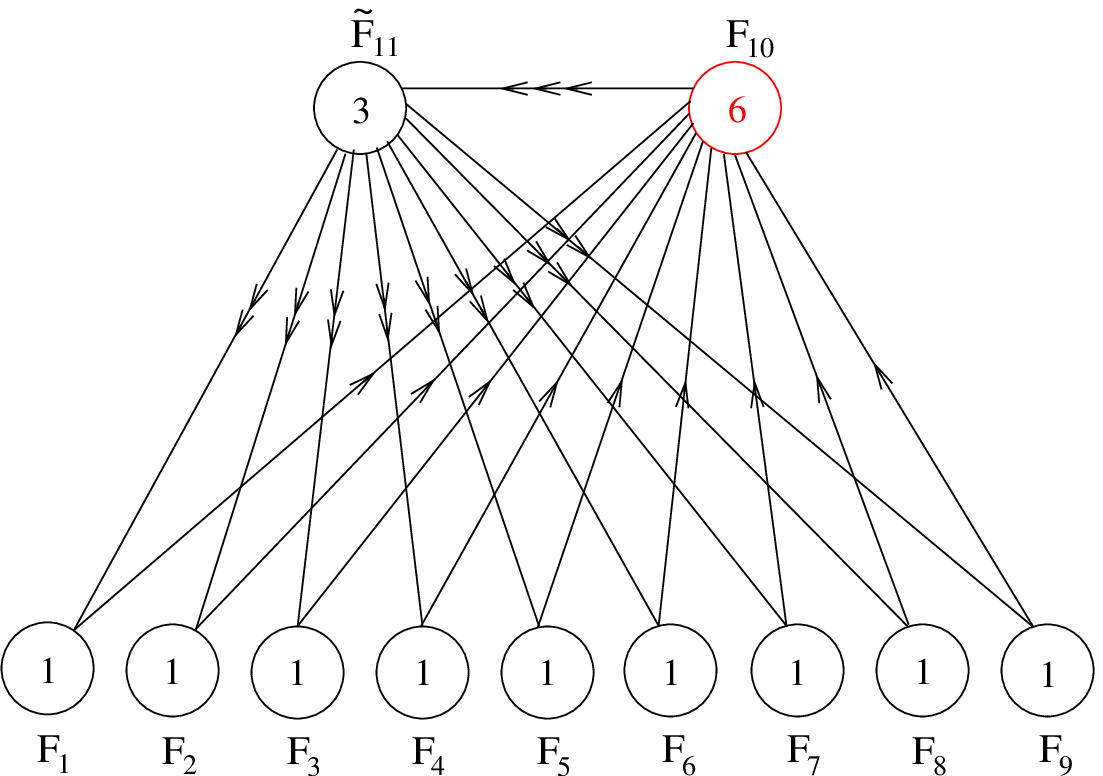}}

Next, it is possible to perform a Seiberg duality on the node $F_{10}$, as in \Verlinde.
This modifies the quiver gauge theory as follows: $F_{11}$ is
replaced by a node  $\tilde F_{11}$ with
\eqn\newel{\ch({\tilde F}_{11}) = (3,\sum_{i=5}^{8}e_i,-2)}
and the multiplicities  at the nodes (for the toy MSSM) are now given by
\eqn\multi{
{\tilde n}_i=1,~i=1,\ldots,9,~~{\tilde n}_{10}=-6,~~{\tilde n}_{11}=3.
}
The spectrum of bifundamentals is now determined by
\eqn\bifundsi{
\chi(F_{10},F_i) = 1,~~\chi({\tilde F}_{11},F_i) = 2 ~{\rm for}~i=1,
\ldots,9,~~\chi({\tilde F}_{11},F_{10}) = 3.
}
The end result is a quiver of the form given in fig. 6 of \Verlinde\
which we reproduce here for convenience.

This quiver can easily accommodate the supersymmetry breaking
model of \quiverDSB.
For instance, the multiplicities
\eqn\susybm{n_{1} = 3M,~n_{10} = 2M,~n_{11} = M}
with all other $n_i$ vanishing, yield the family of theories of interest.
Replacing $n_1$ with any of the $n_{2,\ldots,9}$ would work equally well.
We would like to avoid proliferation of messengers, so for practical
purposes the most interesting case is $M=1$.

\subsec{DSB from the $dP_5$ quiver}

We are also interested in a compact geometry containing a collapsing del Pezzo $dP_5$ surface $S$.
We can determine the corresponding quiver gauge theory starting again with a three-block strongly
exceptional collection on $S$. Such a collection is given by \KN
\eqn\excoldPfive{\eqalign{
&G^a_1=\CO_S(e_4),~~G^a_2=\CO_S(e_5),~~G^b_3=\CO_S(h),~~G^b_4=\CO_S(2h-e_1-e_2-e_3),\cr
&G^c_5=\CO_S(3h-e_1-e_2-e_3-e_4-e_5),~~G^c_6=\CO_S(2h-e_1-e_2),~~G^c_7=\CO_S(2h-e_2-e_3),\cr
&G^c_8=\CO_S(2h-e_1-e_3),
}}
where as before $h$ denotes the hyperplane class and $e_i$, $i=1,\ldots,5$ are the exceptional
curve classes. Then, the associated quiver gauge theory is described by the following diagram

\ifig\sing{The quiver gauge theory associated with a collapsing $dP_5$ surface.}{\epsfxsize2.1in\epsfbox{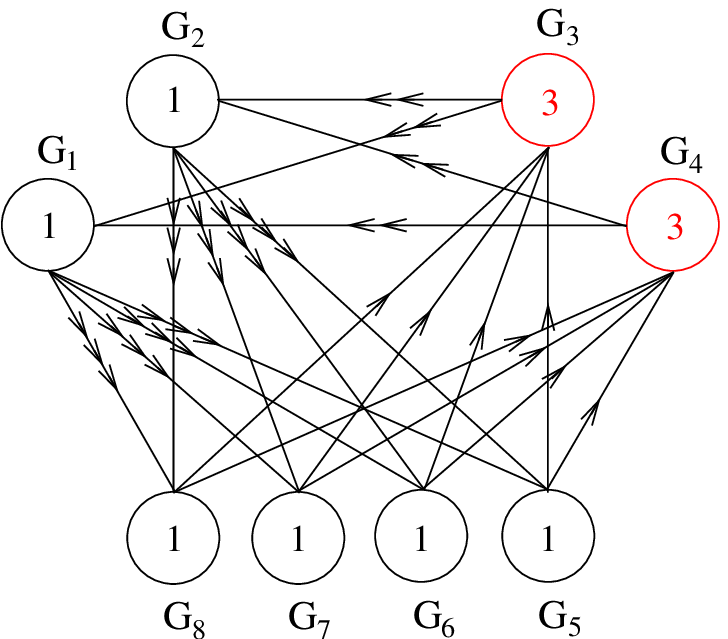}}

It is easy to see that $dP_5$ quiver gauge theory also accommodates the SUSY breaking model of \quiverDSB. One possible
choice of multiplicities is as follows:
\eqn\sussybmi{n_{1}=M,~n_3=3M,~n_5=2M}
with all the other $n_i$ vanishing. Again, the most interesting case is $M=1$.

\newsec{Fractional Brane Supersymmetry Breaking and GUT Models}

Motivated by the previous discussion, in this section we construct examples of IIB
Calabi-Yau orientifolds including fractional branes at del Pezzo singularities.
Since we will focus on examples with fixed O7-planes, our models can also
be regarded as limiting cases of F-theory compactifications  \refs{\SenI,\SenII}.

Consider a IIB compactification on a Calabi-Yau threefold $Z$ equipped with a
holomorphic involution $\sigma : Z\to Z$ which flips the sign of the global holomorphic
three-form
$$ \sigma^*\Omega_Z = - \Omega_Z.$$
Furthermore, let us assume that the fixed locus of $\sigma$ is a smooth complex surface
$R\subset Z$ so that the quotient $Z/\sigma$ is a smooth threefold $P$ and the projection map
$\rho :Z\to P$ is a double cover with ramification divisor $R$.
We will denote by $B\subset P$ the branch
divisor of the double cover.

We construct an orientifold theory by gauging the discrete symmetry
$(-1)^{F_L}\Omega\sigma$
where $\Omega$ is world-sheet parity. According to \refs{\SenI,\SenII}, the resulting
model is equivalent to a F-theory compactification on $X=(Z\times T^2)/\IZ_2$, where $\IZ_2$ acts
as $\sigma$ on $Z$ and simultaneously as $(-1)$ on $T^2$.
It is easy to check that $X$ is an elliptically fibered Calabi-Yau fourfold over $P$ with
$D_4$ singular fiber along the branch divisor $B$. This elliptic fibration admits complex structure
deformations which will modify the singular fibers and the discriminant. Such deformations correspond
to more general F-theory compactifications away from the orientifold limit.

In the following we will be interested in models in which the base $P$ develops collapsing
del Pezzo singularities away from the branch locus of $p:Z\to B$. More precisely, we would like
to find a collection $S_1,\ldots, S_k$ of del Pezzo surfaces on $P$ which do not
meet the branch locus $B$, nor each other, and a map
$p:P\to {\widehat P}$ which contracts $S_1,\ldots, S_k$ to singular points $p_1,\ldots, p_k$
on ${\widehat P}$.

Assuming that these conditions are met, note that the inverse image of each surface $S_i$
via the double cover $\rho :Z\to P$ is a pair of disjoint del Pezzo surfaces $S_i', S_i''$
in $Z$. The involution $\sigma:Z\to Z$ maps $S_i'$ isomorphically to $S''_i$.
Moreover, the contraction map $p$ projects the branch locus $B$ onto a smooth divisor
${\widehat B}\subset {\widehat P}$ which is isomorphic to $B$ and supported away from the
singular points of ${\widehat P}$. Let ${\hat \rho}:{\widehat Z} \to {\widehat P}$ be the double
cover of ${\widehat P}$ branched along ${\widehat B}$. Then $Z$ is isomorphic to the fiber product
${\widehat Z} \times_{{\hat \rho}} P$ and we have a commutative diagram
\eqn\commdiag{
\xymatrix{
Z \ar[r]^\rho \ar[d]^z & P \ar[d]^{p} \\
{\widehat Z} \ar [r]^{\hat \rho} & {\widehat P} \\
}}
The map $z:Z\to {\widehat Z}$ contracts $S_i',S_i''$ to singular points $p_i',p_i''$ on ${\widehat Z}$
which project to the singular points $p_i \in {\widehat P}$ under the map ${\hat \rho}$.

Since $S_i$ are disjoint from $B$, it follows that the infinitesimal neighborhood of $S_i$ in $P$ is
isomorphic to the infinitesimal neighborhood of $S_i'$ (or, equivalently, $S_i''$) in $Z$. Therefore
the $S_i$ must be locally Calabi-Yau surfaces on $P$, even though $P$ is not a Calabi-Yau manifold.
In particular, the normal  bundle to $S_i$ in $P$ must be isomorphic to the canonical
bundle $K_{S_i}$, and the restriction $c_1(P)|_{S_i}$ of the first Chern class of $P$ to
$S_i$ is trivial.

Under these circumstances, the local
physics at the singularities of ${\widehat P}$ is identical to the local physics of
typical del Pezzo singularities in  Calabi-Yau threefolds.
In particular we can introduce fractional branes wrapping collapsing cycles in $S_i$
and study their dynamics as if $P$ were globally Calabi-Yau.
Since the local physics is not sensitive to complex structure deformations
which preserve the singularities, this conclusion will continue to hold when we deform
$X$ away from the orientifold limit.

As explained in the previous section, dynamical supersymmetry breaking can be realized if we place
certain configurations of fractional branes at del Pezzo singularities in Calabi-Yau threefolds.
More specifically, one has to consider fractional branes which in the large radius limit
correspond to D5-branes wrapping holomorphic cycles in the exceptional del Pezzo surface.
This construction seems to be at odds with our set-up since typically O7 orientifold planes
do not preserve the same fraction of supersymmetry as D5-branes. In fact the orientifold projection
considered above maps a D5-brane wrapping a holomorphic curve $C$
in the del Pezzo surface $S_i'\subset Z$ to a anti-D5-brane wrapping the image curve $\sigma(C)$
in $S''_i$. Obviously, for generic values of the K\"ahler parameters this configuration would
break supersymmetry at tree level.
However, we know that in the singular limit fractional D5-branes preserve the same fraction of
supersymmetry as D3-branes. This is clear from the construction of quiver gauge theories
associated to branes at del Pezzo singularities in which configurations of D3-branes and D5-branes
give rise to supersymmetric field theories. In fact there is nothing mysterious about this phenomenon.
It is well established by now that the fraction of supersymmetry preserved by
holomorphic
D-brane configurations is determined by the phase of the associated central charge \refs{\DFR,\Dcat,\AD}.
As we move in the K\"ahler moduli space, the phases of D5-branes and D3-branes change until they
eventually become aligned along a wall of marginal stability.

Since configurations of O7-planes and D3-branes are supersymmetric, it follows that
along such a wall of marginal stability we can also add
fractional D5-branes at the singularities without
breaking tree level supersymmetry. More precisely, suppose we have
$N_i$ D3-branes and $M_i$ D5-branes at the singularity in ${\widehat Z}$ obtained by
collapsing $S_i'$.

The orientifold projection will map this configuration to a system of $N_i$ D3-branes and
$M_i$ anti-D5-branes at the conjugate singularity obtained by collapsing $S''_i$.
The central charges of the two configurations are
\eqn\centralch{
Z_i' = N_i Z_{D3} + M_i Z_{D5},\qquad
Z_i'' = N_i Z_{D3} - M_i Z_{D5},
}
since orientifold projection flips the sign of the D5-brane central charge and preserves the D3-brane
central charge. Since $Z_{D3}$ and $Z_{D5}$ are collinear, it follows that $Z_i,Z_i''$ will also be
collinear as long as the number $N_i$ of D3-branes is sufficiently large.

As explained in section four, we will be in fact interested
in a special point on the marginal stability wall where the low energy effective
theory of the fractional brane system is a quiver gauge theory. This is the point where
the central charges of all objects of the exceptional collection on $S$ are collinear.
In the following we will refer to this special point in the \kah\ moduli space as the quiver point.

Note that tree level supersymmetry is easily achieved in typical F-theory compactifications.
At the quiver point, all
fractional branes corresponding to the exceptional collection
preserve the same supersymmetry as the D3-brane, and they are permuted
by monodromy transformations in the complexified K\"ahler moduli space. Therefore their
central charges must be all equal and aligned to the central charge of the D3-brane.
Since the sum of all fractional branes is a D3-brane, their central charges
must be equal to ${1\over p} Z_{D3}$, where $p$ is the total number of fractional
branes. Fractional D5-branes are linear combinations of some number $q<p$ of fractional branes.
Therefore, $Z_{D5}={q\over p}Z_{D3}$, with $0< q/p<1$.
Therefore the above conditions are easily satisfied in F-theory models with a
large tadpole for three-brane charge.
This number is controlled by ${\chi \over 24}$
where $\chi$ is the Euler character of
the fourfold \SVW,
and in simple examples ${\chi \over 24} \sim 10^2-10^3$. We will be able to compute $\chi$ in one of
our examples and confirm that the above assertion is justified.

It is worth noting that similar configuration of branes have been considered before in toroidal
orientifold models \refs{\BLT,\CU,\LP,\FI,\MS,\CL}.
In that case one typically considers D9-$\overline{\hbox{D9}}$ pairs with magnetic fluxes
in orientifold theories with O3/O7 planes. Such configurations are related by T-duality to D6-branes
intersecting at angles, and preserve tree level supersymmetry
for special values of the K\"ahler parameters. Here we encounter the Calabi-Yau counterpart of this
construction. Other applications of magnetized branes on Calabi-Yau manifolds will be discussed
elsewhere \refs{\DGS}.

\subsec{Concrete examples}

Let us now present a concrete class of models. In addition to the fractional branes
which cause DSB, we would also like our models to exhibit a three generation
GUT sector on the background D7-branes.

We take the base $P$ to be the projective bundle
$\IP\left(\CO_S\oplus K_S\right)$ where $S=dP_k$ is a del Pezzo surface.
Note that $P$ has two canonical sections $S_0, S_\infty$
with normal bundles
\eqn\normal{
N_{S_0/P} \simeq  K_S, \qquad N_{S_\infty/P}\simeq -K_S.
}
The canonical class of $P$ is
$$
K_P = -2S_\infty.
$$
Pick the branch locus $B$ to be a generic smooth divisor in the linear system
$|-2K_P|=|4S_\infty|$. Note that $B$ does not intersect $S_0$, since $S_\infty$
and $S_0$ are disjoint. Then the double cover $\rho:Z\to P$ of $P$ branched along $B$ is a
smooth Calabi-Yau threefold containing two disjoint surfaces $S_0',S_0''$ isomorphic to
$dP_k$, which cover $S_0$.

Note that the section $S_0$ is locally Calabi-Yau in $P$ according to
\normal.\ Moreover, for $k\leq 5$, $P$ is toric, and one can find a toric contraction map $p:P\to
{\widehat P}$ which contracts $S_0$ to a point.
Then one can complete the diagram \commdiag\ by taking the double cover of the cone
${\widehat P}$ branched along the image ${\widehat B}$ of $B$ through $p$.

Next, we deform the elliptic fibration $\pi :X\to P$ preserving the fibration structure
and the singularity in the base. At generic points in the moduli space, we have a smooth
elliptic fibration $\pi:X\to P$ which can be written in standard Weierstrass form
\eqn\weierstrass{
y^2=x^3-fx-g}
where $f,g$ are sections of $-4K_P$ and respectively $-6K_P$. The discriminant is given by
\eqn\discrim{
\delta = 4f^3 -27 g^2.
}
We will denote with capital letters $F,G,\Delta$ the zero divisors of $f,g,\delta$ on $P$.
Note that $\Delta,F,G$ do not intersect $S_0$, therefore the elliptic fiber is constant
along $S_0$. This will allow us to contract the section $S_0$ on $X$, obtaining a
singular fourfold ${\widehat X}$ with an elliptic curve of local $dP_k$ singularities.

In order to obtain
a GUT gauge group on the F-theory seven branes, we have to choose the complex structure moduli
so that $\Delta$ decomposes into two irreducible components
\eqn\irredcomp{
\Delta = \Delta'+\Delta''.}
where $\Delta'=5\Sigma$, where $\Sigma$ is a section of $P$ linearly equivalent to $S_\infty$.
Moreover, $f,g$ should not vanish identically along $\Sigma$.
Note that $\Sigma$ does not intersect the section $S_0$, but it can be brought arbitrarily close to
$S_0$ by complex structure deformations. This means that the mass of the open strings between the
fractional branes at the $dP_k$ singularity and the GUT D7-branes is controlled by a complex
structure modulus of the fourfold.

The chiral matter of the low energy theory is determined in principle by the open string spectrum between
the GUT D7-branes and the D7-branes wrapping the nodal component $\Delta''$ of the discriminant.
However, at the present stage F-theory techniques are not sufficiently developed for an explicit
computation of the spectrum, or at least of the net number of
generations. This question can be more efficiently
addressed invoking heterotic F-theory duality as in section three.

\subsec{Heterotic duals}

The above F-theory models are dual to heterotic four dimensional compactifications on elliptic
fibrations.  We have reviewed some aspects of the duality map and the spectral cover
construction of heterotic bundles in sections \S 3.9 and \S 3.4.

The dual heterotic models are specified by a smooth  Weierstrass model $Y$ over $S$ and
a background bundle of the form $V\times W$ where $V,W$ are stable $SU(5)$ and
respectively $E_8$ bundles over $Y$.
The $E_8$ bundle $W$ corresponds to the hidden sector, while the $SU(5)$ bundle gives rise to the
GUT sector. As explained in \S 3.9, since the F-theory base is a $\IP^1$
bundle over $Y$ without blow-ups, there are no horizontal heterotic
fivebranes on $Y$.
Note that in the notation of \S 3.9, we now have $\CT \simeq K_S$.
Applying the techniques explained there, it is not hard to check that one can enforce a
split $A_4$ singularity along the section $\Sigma\subset P$, which has normal bundle $N_\Sigma
\simeq K_S^{-1}$. This would correspond to an $SU(5)$ bundle $V$ with a
spectral cover $\CC$ in the linear system $|5\sigma +7\pi^*c_1(S)|$.
However, one can easily check using formulas \bundleD\ that such a bundle can
never yield a three generation spectrum.
For $SU(5)$ bundles, the Chern classes \bundleD\ are
\eqn\sufive{
\eqalign{
& \ch_1(V) = -(a+4)\eta + 5(a+2)c_1(S) + 5c_1(\CM)\cr
& \ch_3(V) = \half\eta(\eta-c_1(S))-\eta c_1(\CM).\cr}}
It suffices to substitute
$\eta=7c_1(S)$ in the second equation in \sufive\ obtaining
$$
\ch_3(V)=21-7c_1(S)\cdot c_1(\CM).
$$
The right hand side of this equation is obviously a multiple of $7$, hence it
can never take the value $\pm 3$.

Then how can we obtain a three generation $SU(5)$ GUT in the low energy effective action?
In order to solve this puzzle, we have to look for three generation models in heterotic
vacua with horizontal fivebranes, that is we have to allow a nontrivial $\Xi$ class.
According to \S 3.9, this means that the F-theory base must be a blow-up
of $P$ along a curve isomorphic to $\Xi$. Let us blow-up $P$ along the same
curve $\Xi$ embedded in the section $\Sigma$.
Then the D7-branes carrying the GUT gauge group will wrap the strict transform
$\wSigma$ of the section $\Sigma$, and the class $\eta$ of the $SU(5)$ bundle
must be corrected to
\eqn\dualmapB{
\eta = 7c_1(S) - \Xi.}
The extra term in the right hand side of \dualmapB\ reflects the change in the normal bundle
of $\Sigma$ under the blow-up, as explained in \S 3.8.

Taking into account this correction to the $\eta$ class, let us try again to
find three generations $SU(5)$ bundles. First we have to choose $\CM$ and $a$
so that $\ch_1(V)=0$. There two obvious choices satisfying this condition
\eqn\solsA{
\eqalign{
& A) \quad a=1,\qquad c_1(\CM) = \eta - 3c_1(S)\cr
& B) \quad a=-4,\qquad c_1(\CM) = 2c_1(S).\cr}}
One can probably find many more solutions, but we will focus only on these two
cases in the following.
Substituting \dualmapB\ and \solsA\ in the formula \sufive\ for $\ch_3(V)$ we obtain
\eqn\solsB{
\eqalign{
& A) \quad \ch_3(V) = -\half(7c_1(S)-\Xi)(2c_1(S)-\Xi) \cr
& B) \quad \ch_3(V) = \half (7c_1(S)-\Xi)(2c_1(S)-\Xi). \cr}}

In addition to the constraint $\ch_3(V)=\pm 3$, we would also like $\Xi$ to be a collection of disjoint
$(-1)$ curves on $S$. According to \S 3.10, this is required in order to
stabilize the sizes of the exceptional divisors of the blow-up $\wP\to P$.
Taking into account all these constraints, we have found two classes of
solutions. Recall that we denote by $h$ the hyperplane class on $S=dP_k$ and by
$e_1,\ldots,e_k$ the exceptional curve classes.

$i)\ Example\ I.$ Take $S=dP_5$ and
$$
\Xi = \Gamma_1+\ldots + \Gamma_5
$$
where $\Gamma_1,\ldots,\Gamma_5$ are disjoint smooth rational curves on
$S$. For example we can take $\Gamma_i=e_i$, $i=1,\ldots,5$.
Then one can check that
$$
(7c_1(S)-\Xi)( 2c_1(S)-\Xi) = 6
$$
which yields $\ch_3(V) = \mp 3$ in the two cases listed in \solsB.\ In this
case we can easily compute the Euler characteristic of the fourfold $X$,
$\chi(X)=4128$. Therefore, the three-brane charge tadpole is large enough such that the
technical condition necessary for the alignment of the D3-brane and D5-brane
charge at the point in the \kah\ moduli space where the negative section $S_0$
collapses to zero size is satisfied.

$ii)\ Example\ II.$ Take $S=dP_8$ and
$$
\Xi = \Gamma_1+\Gamma_2
$$
where $\Gamma_1,\Gamma_2$ are any two disjoint $(-1)$ curves on $S$ such as
$e_1,e_2$. Then we obtain again
$$
(7c_1(S)-\Xi)(2c_1(S)-\Xi) = 6.
$$

In each of the these examples, one can construct the SUSY-breaking quiver
theory described in \S4.

\subsec{Comments on possible runaway behavior in \kah\ moduli space}

In the present construction, say for definiteness Example II above, all
\kah\ moduli of the $dP_8$ surface $S_0$ are present as \kah\ moduli
of the Calabi-Yau threefold. Therefore, one must worry about stabilizing
each of them.\foot{This also indicates that Example II would be an
ideal setting for compact embeddings of the Verlinde-Wijnholt model
\Verlinde.  The presence of all $dP_8$ moduli gives one the
freedom to tune Yukawa and FI terms (at least before accounting for
moduli stabilization), a necessary step in making the model
realistic.  For our present
purposes, we wish to use the cone over $dP_8$
to embed a SUSY breaking sector instead.}
(The constructions of \S6 are advantageous in this respect, since as
we shall see there only a single
linear combination of the \kah\ moduli of the $dP_8$ is
nontrivial in $H^{1,1}$ of the Calabi-Yau space).

There are naively sufficiently many vertical arithmetic genus one divisors for
this task; in the IIB picture, these project
to $\IP^1$ bundles over the relevant curves in the del Pezzo surface.
However, the fractional brane configuration
\susybm\ involves one of the exceptional curves in the del Pezzo surface. It is
therefore uncertain that the relevant instanton contributes.  The volume of this
curve plays the role of a FI parameter $R$.  As described in
e.g. \S3.2.2-3.2.3\ of the second reference in \quiverDSB,
this dynamical FI term may cause the SUSY-breaking vacuum to run
away to infinity.  We briefly review their discussion here, for
completeness, and point out several caveats.

In a limit of the $U(3) \times U(2) \times U(1)$ gauge theory where
the $U(2)$ is most strongly coupled, one can derive as in the
reference above a description of the physics which is governed by
only three light chiral multiplets: call them $M$, $Z$ and
$R$.  $M$ and $Z$ are open string fields, while $R$ is the \kah\
modulus.  The real part of $R$ plays the role of an FI term in
the gauge theory -- by abuse of notation we shall denote this by
$R$ as well.  An $SU(3)$ factor in the (Seiberg dual) gauge theory
generates a nonperturbative superpotential, characterized by
dynamical scale $\Lambda$.
Making the very strong assumption that there are canonical kinetic terms,
and expanding for small $R$, the theory has an F-term potential
\eqn\fterm{V_F \sim  |M|^2 + \Lambda^2 |Z + \Lambda^4 M^{-3/2}|^2}
and a D-term potential
\eqn\dterm{V_D \sim (|Z|^2 - R)^2~.}
We emphasize that the assumption of canonical kinetic terms is far
from justified; in such a theory, one would generically expect the
\kah\ potential to include complicated structure at ${\cal O}(\Lambda)$
in field space. Proper account of this could change the conclusions of
even this heuristic discussion.

With these assumptions, it is easy to see that in the presence of the dynamical
FI term, the theory has an unstable vacuum.  The SUSY breaking vacuum
of \quiverDSB, can relax to a vacuum at $R \to \infty$, $Z \to \infty$
and $M \to 0$.

In the language of the brane configuration, one can gain some intuition
for this phenomenon as follows.  Classically (i.e. at $\Lambda \to 0$)
the brane configuration is BPS at $R=0$.  But in fact, as mirrored in
the field theory, there is a full line in \kah\ moduli space along
which the brane configuration remains supersymmetric.  One can take
$R>0$, $Z \sim \sqrt{R}$ and $M=0$.  The non-perturbative SUSY breaking
at $R=0$ may then try to relax by moving along this (classical) flat
direction, towards the larger volume classically BPS brane configuration.

It is clear from this discussion that the determination of the status
of the SUSY breaking vacuum in the compact model is not amenable to a
simple local analysis.  Even if the full \kah\ potential does not
contain enough structure to meta-stabilize the DSB vacuum at
${\cal O}(\Lambda)$
in field space, further effects in the compactification manifold
which depend on $R$ (notably, \kah\ potential corrections and possible
brane instantons) can prevent the runaway behavior.
This possibility is also suggested by the fact that $R$ cannot become
arbitrarily large while keeping the other \kah\ moduli fixed (this
runs into a wall of the \kah\ cone); but the other \kah\ moduli can be fixed
by brane instanton effects in this model.
Because the models
of \S3\ have no such issue, we will not try to pursue the analysis of
this model in further detail here.  A model based on the same SUSY
breaking quiver is also described in \S6, where however there are
fewer del Pezzo moduli and the
existence of a stable vacuum seems very plausible.

\newsec{Compactifications Incorporating the D3-Brane MSSM}

In this section we construct another class of models which contain an MSSM
sector realized as in \refs{\Verlinde} instead of a GUT sector on D7-branes.

We will use the same kind of geometric set-up encoded in the commutative diagram
\commdiag,\ which is reproduced below for convenience
\eqn\commdiagB{
\xymatrix{
Z \ar[r]^\rho \ar[d]^z & P \ar[d]^{p} \\
{\widehat Z} \ar [r]^{\hat \rho} & {\widehat P}.\\
}}
Recall that $Z$ is a Calabi-Yau threefold equipped with a holomorphic involution $\sigma :Z\to Z$
which fixes a smooth divisor $R$. $P$ is the quotient $Z/\sigma$ and $\rho:Z\to P$ is the canonical
$2:1$ projection map.
Moreover, we assume that we can find a collection  $S_i',S_i''$, $i=1,\ldots,n$ of
del Pezzo surfaces on $Z$
disjoint from $R$ so that $\sigma$ maps $S_i'$ to $S''_i$, $i=1,\ldots,n$.
The del Pezzo surfaces must be contractible on $Z$, and $z:Z\to {\widehat Z}$ is a
contraction map which collapses $S_i', S_i''$ to singular points $p_i',p_i''$ on ${\widehat Z}$.

Suppose we can find such a geometric set-up with $n=2$. Then we can place conjugate D3-D5 systems at the
singular points $(p_1',p_1'')$, obtaining a supersymmetric configuration as explained
in the previous section. We can also place fractional D3-branes at the remaining singular points
$(p_2',p_2'')$ realizing a supersymmetric standard model as in \Verlinde. Open strings stretching between
the two types of D-brane configurations mediate supersymmetry breaking. The mass of these open strings
is controlled by the relative position of the points $(p_1',p_2')$ (or, equivalently $(p_1'',p_2'')$)
which is a complex modulus of $Z$.

Next let us construct a concrete example. We start with a smooth Weierstrass model $Z'$ over a del Pezzo
surface $S=dP_5$. Here we regard $dP_5$ as a
four point blow-up of the Hirzebruch surface $\IF_0=\IP^1\times \IP^1$.
Let $p_i',p_i''$, $i=1,2$ denote the centers of the blow-ups on $\IF_0$, and let
$e_i',e_i''$, $i=1,2$ denote the exceptional curves.
If the fibration is generic, the restriction of the elliptic fibration to any $(-1)$ curve $C$ on $S$
is isomorphic to a rational elliptic surface with $12$ $I_1$ fibers, usually denoted by $dP_9$.
Therefore, taking $C$ to be each of the four exceptional curves, we obtain four rational elliptic
surfaces $D_i',D_i''$. The exceptional curves $e_i',e_i''$ can be naturally identified
with sections of the rational elliptic surface $D'_i,D''_i$. We can also naturally regard them as
$(-1,-1)$ curves on $Z'$ embedding $S$ in $Z'$ via the section of the Weierstrass model.

Threefolds $Z'$ of this form appear quite often in F-theory, where the blow-ups in the base are
associated to point-like small instantons in the dual heterotic string \refs{\MVII,\ATasi,\Apoint,\AM}.
In this context, it is known that models with different number of blow-ups are related by
extremal transitions which proceed as follows. One first performs a flop on the $(-1,-1)$
curves $e_i',e_i''$ in $Z'$, obtaining an elliptic fibration $Z\to \IF_0$ with two complex dimensional
components in the fiber. More precisely, the fibers over the points $p'_i,p''_i$ have two components:
a rational $(-1,-1)$ curve obtained by flopping one of the curves $e_i',e_i''$ and a $dP_8$
 del Pezzo surface.
We will denote the $dP_8$ components by $S_i',S''_i$ as above.
Next, one can contract the del Pezzo surfaces in the fiber, obtaining a singular elliptic fibration
${\widehat Z}$ over
$\IF_0$, which can be eventually smoothed out by complex structure deformations.

Our example will be a threefold $Z$ obtained at the intermediate stage of the extremal transition.
All the required elements are in place apart form the holomorphic involution $\sigma$, which can be
realized as follows. In the above construction, we restrict ourselves to a class of symmetric $dP_5$
surfaces $S$ obtained by blowing-up conjugate points on $\IF_0$ under a holomorphic
involution $\kappa:\IF_0\to \IF_0$. Using homogeneous toric coordinates
\eqn\toric{
\matrix{ & Z_1 & Z_2 & Z_3 & Z_4 \cr
\IC^* & 1 & 1 & 0 & 0 \cr
\IC^* & 0 & 0 & 1 & 1 \cr}}
on $\IF_0$, $\kappa$ is given by
\eqn\baseinv{
\kappa: (Z_1,Z_2,Z_3,Z_4) \to (Z_1,-Z_2, Z_3, Z_4).}
Note that the fixed loci of $\kappa$ are two disjoint curves on $\IF_0$
determined by the equations
\eqn\fixedcurves{
 Z_1=0\qquad Z_2=0.}
We pick $(p_1',p_1')$ and $(p_2',p_2'')$ to be pairwise conjugate under $\kappa$ and away from the
fixed curves. Then a simple local computation shows that $\kappa$ lifts to a holomorphic involution of
$S$ which maps $e_i'$ isomorphically to $e''_i$. Abusing notation, we will also denote by $\kappa$ the
lift to $S$. The distinction should be clear from the context.
Note that the fixed locus of $\kappa$ on $S$ consists of two curves $C_1,C_2$
which are the strict transforms of
\fixedcurves.\

The elliptic fibration $Z'$ can be written as a hypersurface in the projective bundle
$\CP=\IP\left(\CO_S\oplus K_S^{-2} \oplus K_S^{-3}\right)$ over $S$.
The involution $\kappa:S\to S$ constructed in the last paragraph can be trivially lifted to the total space
of $\CP$ by taking the action on the homogeneous coordinates along the fibers to be trivial.
Then we can obtain a symmetric threefold $Z'$ by taking the defining polynomials
of the Weierstrass model to be invariant under the involution $\kappa: S \to S$. Any such threefold
would be preserved by the holomorphic involution on the ambient space $\CP$, therefore it is equipped with
an induced holomorphic involution $\sigma'$. By construction, $\sigma'$ maps a rational elliptic
surface $D_i'$ to a rational elliptic surface $D_i''$, and the induced map is compatible with the fibration
structure.

Moreover, one can check that we also have an induced holomorphic involution $\sigma$ on
the threefold $Z$ obtained from $Z'$ by flopping the curves $(e_i',e_i'')$. In order to see this,
let us recall the geometric description of the flop \refs{\JK}.
One first blows-up $Z'$ along the $(-1,-1)$ curves
$(e_i',e_i'')$ obtaining a non-Calabi-Yau threefold $T$. The exceptional divisors on $T$ are four
surfaces $(Q_i',Q_i'')$, $i=1,2$ isomorphic to $\IF_0$. The involution $\sigma'$ on $Z'$ lifts naturally to
an involution $\tau:T\to T$ which maps $Q_i'$ isomorphically to $Q_i''$.
Each of the surfaces $Q_i', Q_i''$ is equipped with two rulings  which will be denoted by
$a_i',b_i'$, and respectively $a_i'', b_i''$, $i=1,2$. These rulings are preserved by the holomorphic
involution.

Now, there are two possible contractions of
$T$. One can find a contraction map which collapses all the $a$-rulings $a_i',a_i''$ on the surface
$Q_i',Q_i''$, obtaining $Z'$. Alternatively, one can find another contraction map
which collapses the $b$-rulings $b_i',b_i''$, obtaining the threefold $Z$ related to $Z'$ by a flop.
Since the involution $\tau$ preserves the rulings, it follows that in both cases it induces involutions
on $Z'$ and respectively $Z$. In the first case, this is just the holomorphic involution $\sigma'$
we started with. In the second case, we obtain a holomorphic involution $\sigma$ on $Z$ with all
the required properties.

Some final remarks are in order.

$i)$ Note that the fixed point set of $\sigma:Z\to Z$ consists of two K3 surfaces obtained by restricting
the Weierstrass model to the fixed curves $C_1,C_2$ of $\kappa$ in the base $S$.

$ii)$ By construction, the relative position of the points $(p_1',p_2')$ on $\IF_0$ is a complex
modulus of $Z$ which controls the mass of the open strings mediating supersymmetry breaking.

$iii)$ Note that in the present construction the \kah\ moduli of the compact
threefold $Z$ can control only the overall size of the collapsing del Pezzo
surfaces $S_i',S_i''$, $i=1,2$. One cannot control the relative size of the
exceptional curves within each surface.

Point $iii)$ actually means it is unlikely that one can arrange for
a reasonable
symmetry-breaking pattern in the embedding of \Verlinde\ in this compact
model -- attempts to choose FI terms to make the model realistic
will result in hypercharge breaking.
Readers who are interested in compact embeddings of
that theory where the desired Yukawa and FI terms of \Verlinde\ ${\it can}$
be obtained by tuning closed string parameters,
are reminded that the Calabi-Yau fourfold in
Example II of \S5.2\
can be used for this purpose.

\newsec{Conclusion}

Many features of MSSM phenomenology are governed, not so much by the
dynamics of supersymmetry breaking, but rather by the mechanism by
which SUSY breaking is transmitted to the Standard Model. In this
work we have initiated an attempt to construct stringy models where
gravity mediation is subdominant to other mediation mechanisms.
There are many obvious directions for future work:
\smallskip
\noindent$\bullet$
The models of \S3\ have large $\Lambda_H$ and messenger mass $M$, as well as
a large messenger index.  Finding examples with smaller $\Lambda_H$, $M$
and messenger index would clearly be desirable, and may enable construction
of models with all relevant dynamics taking place much closer to the
TeV scale.
\smallskip
\noindent$\bullet$
Several natural questions arise in this setting that could be attacked
with the statistical approach to compactifications advocated in \Dougstat.
For instance, arranging for small messenger mass $M << M_s$ can be
regarded as a tuning of parameters.
Understanding in detail what fraction of flux vacua allow complex structure
stabilization with small $M$ could give one some sense of the difficulty
of arranging for gauge mediation to be the dominant mechanism of
supersymmetry breaking transmission, even in cases where hidden sector
DSB dominates over flux-induced moduli F-terms.
A discussion of several other relevant questions one could pursue
in the statistical framework appears in \DGT.
\smallskip
\noindent$\bullet$
There
are several issues surrounding the compactification of models using the
SUSY BOG mechanism of SUSY breaking \quiverDSB\ that need to be clarified.
The field theory itself may or may not admit metastable vacua at
${\cal O}(\Lambda)$ in field space once the FI terms are rendered
dynamical; and global effects may stabilize the FI terms in any case.
Precise circumstances in which the various possibilities occur should be
specified.
\smallskip
\noindent$\bullet$
The DSB quiver theories which have been discovered so far \quiverDSB,
arise on
D-branes at singularities of Calabi-Yau threefolds.
There are singularities of F-theory fourfold compactifications where one can
imagine
fractional branes localized at collapsed cycles, and the normal
bundle of the brane in the base of the elliptic fibration, is
such that the configuration could never arise in a Calabi-Yau threefold.
It is quite plausible that more interesting examples of DSB from
fractional branes in F-theory, will arise by analyzing the physics
of these singularities.
\smallskip
\noindent$\bullet$
We have focused on gauge mediation as a possible solution to the flavor
problems of SUSY breaking in string theory.  From a conceptual perspective,
since many models of gauge mediation work at parametrically low energies
and do not require assumptions about the UV embedding, it is less urgent
to UV complete gauge mediated models than various alternatives (perhaps
the most urgent issue being the need to find a plausible solution to
the $\mu$ problem).
Finding stringy avatars of gaugino mediation \GM\ and anomaly mediation
\Anom\ (including some solution to the problem of tachyonic sleptons in
the latter case) would certainly be worthwhile.
\smallskip
\noindent$\bullet$ Due to the UV insensitivity of gauge mediation, it
would make sense (somewhat in the spirit of \Verlinde) to engineer
non-compact brane models containing an MSSM or GUT, a SUSY breaking
sector, and the messengers.  The resulting gauge mediated spectrum
would be largely insensitive to the global details of compactification
in any compact embedding.  While one can obviously infer some such models
by taking limits of our compact constructions, it may be easier to
identify promising classes of models in the non-compact setting first.
Early work in this direction appears in \Brodie.
\smallskip
\noindent$\bullet$ Finally, the UV properties of string theory may
suggest novel new possibilities for the transmission of
SUSY breaking, or realizations
of gravity mediation that solve the flavor problem.  Developments
along these lines would be very interesting.
\vfill\eject

\centerline{\bf{Acknowledgements}}
\medskip
We would like to thank S. Dimopoulos, M. Dine, A. Grassi,
R. Kitano, J. McGreevy, T. Pantev, R. Reinbacher, N. Seiberg,
R. Sundrum, H. Verlinde and J. Wacker for interesting
discussions.  
We also thank D. Berenstein, K. Intriligator and N. Seiberg for
helpful correspondence.
S.K. enjoyed the hospitality of the
Amsterdam string theory group during the completion of this work.
S.K. was supported in part by a David and Lucile Packard Foundation
Fellowship for Science and Engineering. B.F., S.K. and P.S. were
also supported by the NSF under grant PHY-0244728 and the DOE under
contract DE-AC03-76SF00515.

\listrefs
\end